\documentclass[longauth]{aa} 

\usepackage{graphicx}
\usepackage{natbib}
\usepackage{flexisym}

\usepackage{letltxmacro}
\usepackage{longtable}
\usepackage{amsmath}
\usepackage{xcolor}
\LetLtxMacro\FlexiSymTextPrime\textprime
\DeclareTextCommandDefault{\textprime}{\FlexiSymTextPrime}  
\usepackage{subcaption}  
\usepackage{float}     

\usepackage{graphicx}
\usepackage{caption}
\usepackage{subcaption}

\usepackage{longtable}
\usepackage{placeins} 

\usepackage{longtable}
\usepackage{booktabs}
\usepackage{pgfplotstable}
\usepackage{siunitx}

\usepackage{hyperref}
\hypersetup{colorlinks=True, linkcolor=blue, citecolor=blue}

\usepackage{txfonts,textcomp}

\begin{document}

   \title{Resolving the extended interstellar medium of infrared galaxies} \subtitle{I. Polycyclic aromatic hydrocarbon emission}

    \author{C. M. Lofaro\inst{1,2}, T. Díaz-Santos\inst{2,3}, L. Armus\inst{4}, V. Charmandaris\inst{1,2,3}, T. Lai\inst{4}, M. Pereira-Santaella\inst{5}, J. McKinney\inst{6}, R. Fernández-Aranda\inst{1,2,7}, G. Donnelly\inst{8}, M. Bianchin\inst{9,10}, Y. Song\inst{11,12}, I. Shivaei\inst{7}, L. A. Boogaard\inst{13}, G. Rodighiero\inst{14,15}, S. Linden\inst{16}, G. C. Privon\inst{17,18,19}, C. Ricci\inst{20,21}, L. Lenkić\inst{4}, H. Inami\inst{22}, S. Stierwalt\inst{23}, V. U\inst{4,24}, D. Kakkad\inst{25}, P. Van der Werf\inst{13}, M. Sánchez-García\inst{2}, A. Evans\inst{17,19}, and D. B. Sanders\inst{26}}
    
    \institute{
    Department of Physics, University of Crete, 70013 Heraklion, Greece\\
    \email{clofaro@ia.forth.gr}
    \and
    Institute of Astrophysics, Foundation for Research and Technology--Hellas (FORTH), Heraklion, GR-70013, Greece
    \and
    School of Sciences, European University Cyprus, Diogenes street, Engomi, 1516 Nicosia, Cyprus
    \and
    IPAC, California Institute of Technology, 1200 E. California Boulevard, Pasadena, CA 91125, USA
    \and
    Instituto de Física Fundamental, CSIC, Calle Serrano 123, 28006 Madrid, Spain
    \and
    Department of Astronomy, The University of Texas at Austin, Austin, TX, USA
    \and
    Centro de Astrobiología (CAB), CSIC-INTA, Carretera de Ajalvir km 4, Torrejón de Ardoz, E-28850, Madrid, Spain
    \and
    Ritter Astrophysical Research Center and Department of Physics and Astronomy, University of Toledo, Toledo, OH 43606, USA
    \and
    Instituto de Astrofisica de Canarias, Calle Via Lactea, s/n, E-38205, La Laguna, Tenerife, Spain
    \and
    Departamento de Astrofisica, Universidad de La Laguna, E-38206, La Laguna, Tenerife, Spain
    \and
    European Southern Observatory, Alonso de Córdova, 3107, Vitacura, Santiago, 763-0355, Chile
    \and
    Max-Planck-Institut f\"{u}r Radioastronomie, Auf dem H\"{u}gel 69, 53121, Bonn, Germany
    \and
    Leiden Observatory, Leiden University, PO Box 9513, 2300 RA Leiden, The Netherlands
    \and
    Dipartimento di Fisica e Astronomia, Università di Padova, Vicolo dell’Osservatorio 3, 35122 Padova, Italy
    \and
    INAF-Osservatorio Astronomico di Padova, Vicolo dell’Osservatorio 5, 35122 Padova, Italy
    \and
    Steward Observatory, University of Arizona, 933 North Cherry Avenue, Tucson, AZ 85721, USA
    \and
    National Radio Astronomy Observatory, 520 Edgemont Road, Charlottesville, VA 22903, USA
    \and
    Department of Astronomy, University of Florida, P.O. Box 112055, Gainesville, FL 32611, USA
    \and
    Department of Astronomy, University of Virginia, 530 McCormick Road, Charlottesville, VA 22903, USA
    \and
    Department of Astronomy, University of Geneva, ch. d'Ecogia 16, 1290, Versoix, Switzerland
    \and
    Instituto de Estudios Astrofisicos, Facultad de Ingenieria y Ciencias, Universidad Diego Portales, Av. Ejercito Libertador 441, Santiago, Chile
    \and
    Hiroshima Astrophysical Science Center, Hiroshima University, 1-3-1 Kagamiyama, Higashi-Hiroshima, Hiroshima 739-8526, Japan
    \and
    Occidental College, Physics Department, 1600 Campus Road, Los Angeles, CA 90042, USA
    \and
    Department of Physics and Astronomy, 4129 Frederick Reines Hall, University of California, Irvine, CA 92697, USA
    \and
    Centre for Astrophysics Research, Department of Physics, Astronomy and Mathematics, University of Hertfordshire, Hatfield, AL10 9AB, UK
    \and
    Institute for Astronomy, University of Hawaii, 2680 Woodlawn Drive, Honolulu, HI 96822, USA
    }

  \date{Received 7 February 2026 / Accepted 15 May 2026 }
   
   \titlerunning{PAHs in GOALS}
   \authorrunning{Lofaro C. M. et al.}

\abstract
   {Luminous infrared galaxies (LIRGs) represent a key phase of galaxy evolution that is characterized by intense star formation, frequent mergers, and in some cases significant active galactic nucleus (AGN) activity. Polycyclic aromatic hydrocarbons (PAHs) and far-infrared (FIR) cooling lines provide powerful diagnostics of the physical conditions of the interstellar medium (ISM) and the efficiency of photoelectric heating of the gas, but most previous studies have relied on nuclear or galaxy-integrated measurements.}
   {We aim to investigate how PAH properties and photoelectric heating efficiency vary spatially across LIRGs, and to assess the relative roles of star formation and AGN activity in regulating gas heating and dust grain processing on kiloparsec scales.}
   {We used spatially resolved mid-infrared (MIR) spectroscopy from \textit{Spitzer}/IRS and FIR line maps from \textit{Herschel}/PACS for a sample of 67 nearby LIRGs from the Great Observatories All-sky LIRG Survey (GOALS) sample. Spectra were extracted in concentric annuli and matched square-aperture grids to probe regions from the nuclei out to $\sim$14\,kpc. We measured MIR PAH features and emission lines using the \textsc{CAFE} spectral decomposition tool, while FIR emission lines were measured through Gaussian profile fitting. We then traced the photoelectric heating efficiency by combining PAH emission with FIR cooling lines after isolating the neutral-gas contribution.}
   {In starburst galaxies, the PAH~11.3/7.7 ratio increases from the nuclear regions out to $\sim$5~kpc, while the PAH~6.2/7.7 ratio remains approximately constant, indicating neutral and large PAH populations. At intermediate galactocentric distances (5--8~kpc), the PAH~11.3/7.7 ratio continues to rise as PAH sizes begin to shift toward smaller grains, marking a transition regime. Beyond $\sim$8~kpc, the PAH~11.3/7.7 ratio decreases and the PAH~6.2/7.7 ratio increases, signaling increasingly ionized and smaller PAHs, which is consistent with a photo-destruction scenario driven by a strong radiation field close to the central starburst. Nuclear and inner extended regions in strong AGN systems are dominated by more neutral PAH populations and harder radiation fields, whereas at larger radii (> 5~kpc) PAHs become increasingly ionized and converge toward the starburst (SB) locus with a declining radiation field intensity, which indicates the influence of the AGN is diminishing. Our results also show that powerful AGNs can influence the average PAH ionization states at large radii (6--14~kpc) with the same effectiveness that moderate AGNs achieve at smaller scales (1--6~kpc). The photoelectric efficiency increases with galactocentric distance in both SB and AGN-hosting galaxies, due to the decreasing intensity of the radiation field at progressively larger radii. In SB, the increase in photoelectric efficiency is accompanied by a progressive shift of the PAH population toward more neutral charge states, whereas in AGN-dominated systems this correlation appears to be inverted.}
   {Our results demonstrate that both star formation and AGN activity regulate PAH properties over kiloparsec scales (up to ${\sim}$5~kpc) in LIRGs, with the influence of the AGN weakening with galactocentric distance. This work provides the first quantitative, spatially resolved constraints on the kiloparsec-scale extent over which AGN activity impacts MIR dust grain properties in luminous dusty galaxies as a population.}


   \maketitle

\section{Introduction}
One of the most energetic phenomena in the Universe occur in extreme systems known as luminous infrared galaxies (LIRGs; $L_{{IR[8-1000\mu{m}]}}> 10^{11}\,L_{\odot}$) and ultraluminous infrared galaxies (ULIRGs; $L_{{IR[8\text{--}1000\,\mu{m}]}}> 10^{12}\,L_{\odot}$; see \citealt{Sanders96} for a review), whose high infrared emission arises from the reprocessing by dust of intense star formation and, in some cases, active galactic nuclei (AGNs) activity (e.g., \citealt{Petric11, Alonso-Herrero12}). 

Luminous infrared galaxies span the full evolutionary merger sequence, from isolated galaxies to advanced mergers. They are particularly significant among the star-forming galaxy population as they exhibit elevated star formation rates (SFRs) and specific SFRs (sSFRs), resulting from the concentration of large quantities of gas and dust toward their nuclei \citep{Sanders96}, which, in interacting systems, is driven by the loss of angular momentum during the dynamical interaction \citep{Hopkins08}. This phase profoundly impacts the evolution of merging galaxies, triggering a morphological transformation into an elliptical galaxy after the suppression of star formation and black hole growth due to feedback from the dominant energy source(s) (e.g., \citealt{Hopkins08}). Therefore, LIRGs represent a critical phase in galaxy evolution studies, and potentially highlight gaps in our understanding of the transitional processes that shape galaxy morphology and activity.

While LIRGs are uncommon in the local Universe and ULIRGs even rarer, their numbers increase dramatically (by a factor of {$\sim$}100) at z{$\sim$}1 \citep{LeFloc'h05, Magnelli09}. At this redshift, LIRGs dominate the star formation activity and the IR background, whereas at z{$\sim$}2, the so-called cosmic noon, ULIRGs become predominant \citep{LeFloc'h05, Perez-Gonzalez05, Caputi07, Berta11, Magnelli11, Murphy11}. 
This dramatic evolution highlights the importance of understanding the mechanisms that give birth to the LIRG phase. Nearby LIRGs serve as an essential link between normal star-forming galaxies, such as the Milky Way, and the most extreme systems dominated by AGNs (\citealt{Sanders96}).

Among the many infrared tracers of the interstellar medium (ISM) in LIRGs, polycyclic aromatic hydrocarbons (PAHs) are of particular importance. PAHs are small carbonaceous grains that efficiently absorb ultraviolet (UV) photons and re-emit the energy through a series of prominent vibrational features between roughly 3 and 17 $\mu$m \citep{Leger84, Allamandola85, Allamandola89, Draine01}. In addition to their role as tracers of star formation, PAHs play a fundamental role in regulating the thermal balance of the ISM \citep{Galliano08, Tielens08}. PAH emission probes photoelectric heating in photodissociation regions (PDRs), predominantly composed of cold neutral gas \citep{Wolfire03}, which are the interfaces between ionized gas in H{\sc ii} regions and the molecular gas in H$_2$ clouds \citep{Wolfire90}.
When a UV photon is absorbed by a PAH grain, part of its energy can be used to eject an electron via the photoelectric effect. The resulting photoelectron transfers kinetic energy to the surrounding gas through collisions, primarily with H and H$_2$. This heating increases the gas temperature and enables the collisional excitation of C$^+$ and O, which subsequently radiate and act as the dominant cooling channels of the ISM (\citealt{Tielens85, Hollenbach97, Kaufman99} and references therein).
Therefore, the efficiency of photoelectric heating depends sensitively on the physical properties of the dust grains, particularly their size distribution and charge state \citep{Bakes94}. 

Most of the observational evidence for these processes has been drawn almost exclusively from nuclear and/or integrated measurements of galaxies \citep{Smith07}, but spatially resolved \textit{Spitzer}/Infrared Spectrograph (IRS)spectral mapping has demonstrated that mid-infrared (MIR) emission extends well beyond the nucleus. PAH and low-ionization neon emission lines are detected over kiloparsec scales and trace the distribution of the radiation field across the disks of LIRGs, which may include contributions from both local and nearby star-forming regions \citep{Pereira-Santaella10}.

With the advent of JWST, unprecedented ISM studies have been enabled through high-resolution spectroscopy, revealing that PAH properties vary strongly with environment \citep{Linden24} and show systematic changes in AGN-hosting systems \citep{Lai22, Lai23}. These capabilities are now being extended to earlier cosmic epochs: JWST is opening the door to analogous PAH studies at cosmic noon through the PAHSPECS program (GO~5279; P.I. Irene Shivaei), which will characterize both integrated and spatially resolved PAH properties in star-forming galaxies at $z{\sim}1.1$ (Lofaro et al., in prep.; Donnan et al., in prep.; Shivaei et al., in prep.). However, JWST’s limited field of view constrains studies of large-scale feedback effects, which makes complementary wide-field observations essential for probing extended ISM properties.

In this work we investigate the structure and physical characteristics of the spatially resolved PAHs in a subsample of nearby LIRGs from the Great Observatories All-sky LIRG Survey  (GOALS, \citealt{Armus09}) in order to determine whether and how gas heating and cooling are regulated across these extreme environments. We focus on regions at distances from the nuclei larger than what a single pointing of any JWST integral field unit (IFU) can reach. We place particular emphasis on the impact of AGN feedback and the efficiency of the photoelectric effect in different galactic environments. By combining \textit{Spitzer} and \textit{Herschel} observations, we provide the first MIR+far-infrared (FIR) resolved analysis of LIRGs at kiloparsec scales. 

This work is the first of a two-part series aimed at characterizing the extended ISM of LIRGs using a subsample of the GOALS galaxies \citep{Armus09}. 
The present paper (Paper~I) focuses on PAH emission, while the companion paper (Paper~II) investigates the fine-structure emission lines. 

This work is organized as follows. Section~\ref{Goals sample} describes the sample selection and the observations used in this study. In Sect.~\ref{section3} we present the data processing steps, including spectral extraction and emission-line fitting procedures. In Sect.~\ref{results} we investigate the spatially resolved PAH properties and in Sect.~\ref{photo-effiency} we explore the photoelectric efficiency in connection with dust grain characteristics. We summarize our conclusions in Sect.~\ref{conclusions}.
Throughout this work, we assume a flat $\Lambda$ cold dark matter cosmology with $H_0 = 67.4$ km s$^{-1}$ Mpc$^{-1}$, $\Omega_{\rm m} = 0.315$, and $\Omega_{\Lambda} = 0.685$ \citep{Plank20}.

\vspace{-0.2cm}

\section{Sample and observations}
\label{Goals sample}
\color{black}

\subsection{The GOALS sample}
The GOALS sample \citep{Armus09} integrates data from NASA's \textit{Spitzer}, \textit{Chandra}, \textit{Hubble}, and GALEX observatories, along with ground-based observations, to create a detailed imaging and spectroscopic survey of over 200 (U)LIRGs (180 LIRGs and 22 ULIRGs) at low redshift ($z < 0.088$). These (U)LIRGs are a complete, flux-limited sample subset of the IRAS Revised Bright Galaxy Sample (RBGS), which includes 629 extragalactic objects with 60 $\mu$m flux densities greater than 5.24 Jy and Galactic latitudes above five degrees. The LIRGs in GOALS cover the entire range of nuclear spectral types, as defined by traditional optical line-ratio diagrams (including type-1 and type-2 AGNs, LINERs, and starbursts, \citealt{Armus09}), as well as various stages of interaction (major mergers, minor mergers, and isolated galaxies, \citealt{Sabrina13}). 

\begin{figure*}[t] 
    \centering
    \includegraphics[width=0.85\textwidth]{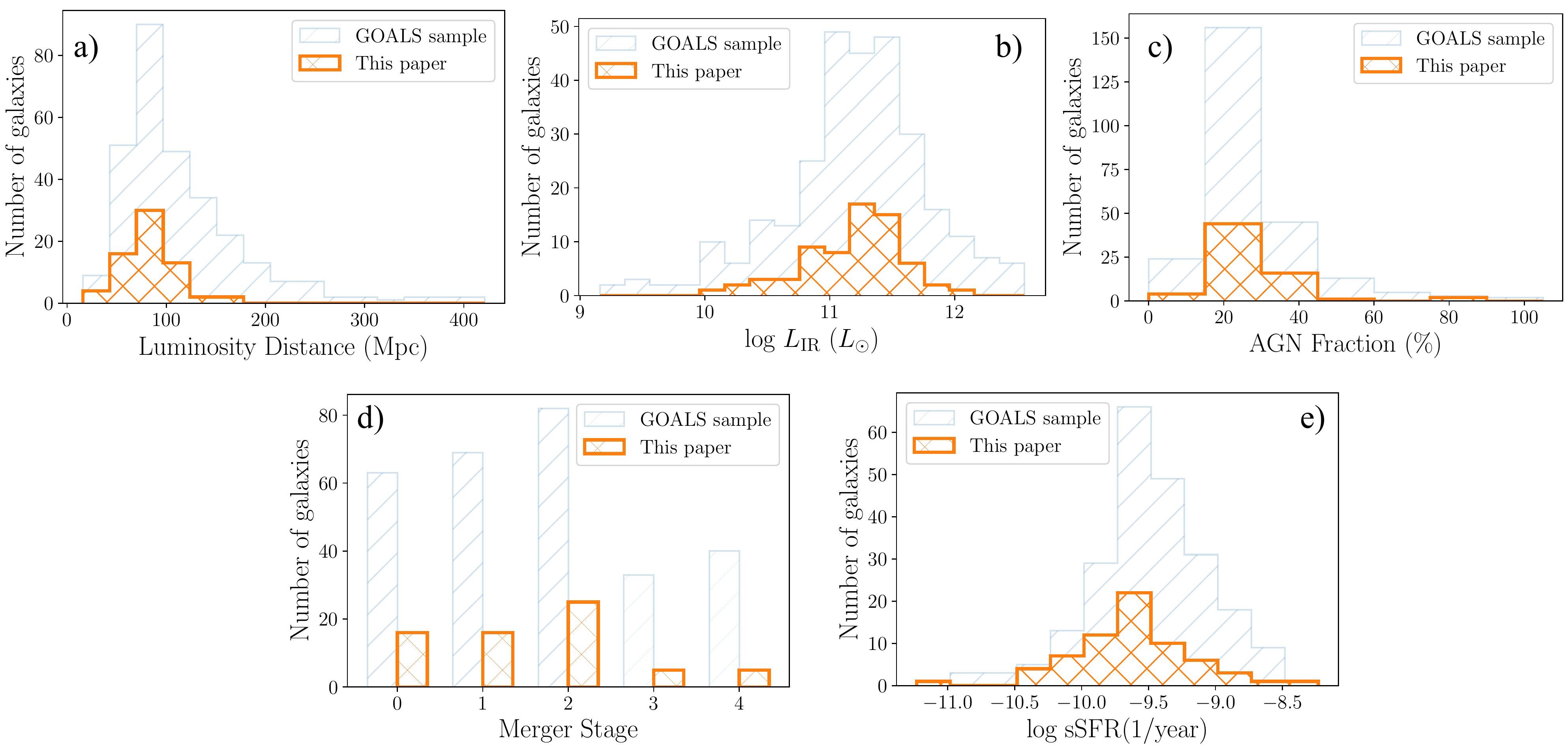}
    \vspace{-0.2cm}
    \caption{Integrated properties of the galaxies of this work compared to the GOALS sample: (a) luminosity distance (D$_{L}$), (b) infrared luminosity (L$_{IR}$), (c) mid-IR f$_{AGN}$, (d) sSFR, and (e) merger stage (see Table~\ref{tab:goals_sample} for more details). Our subsample is representative of the parent sample.}
    \label{fig:histograms}
\end{figure*}

For this study, we selected sources from the GOALS sample based on the availability of spectral mapping observations from both \textit{Spitzer}/IRS and \textit{Herschel}/PACS. Sixty-six galaxies from the GOALS sample meet this requirement. We also included one additional source (NGC 5395) which lacks \textit{Herschel}/PACS spectral maps. The galaxy hosts a prominent AGN and was retained to improve the AGN representation in our otherwise starburst-dominated subsample, enabling more robust comparisons between AGN and starburst properties. An additional selection criterion was the availability of measurements of the fractional contribution of AGNs to the MIR emission of galaxies (f$_{AGN}$), estimated following the methodology of \citet{Veilleux09} and as detailed in \citet{Tanio17}. This approach uses up to five \textit{Spitzer}/IRS diagnostics, depending on data availability: the [NeV]$_{14.3}$/[NeII]$_{12.8}$ and [OIV]$_{25.9}$/[NeII]$_{12.8}$ line ratios, the 6.2 $\mu$m PAH equivalent width \citep{Armus07}, the $S_{30}/S_{15}$ continuum slope (see \citealt{Veilleux09}), and the Laurent diagram \citep{Laurent00}. These indicators quantify the contribution of AGNs to the MIR emission through enhanced ionization and/or dust heating, providing an estimate of the AGN’s fractional contribution to the MIR luminosity ($\alpha^{MIR}_{AGN}$). Their combination yields an average value, $\textlangle \alpha^{MIR}_{AGN} \textrangle$, as described in \citet{Tanio17} that we hereafter refer to as f$_{AGN}$.
Therefore, the final sample for this study consists of 67 galaxies, including 13 systems with two members. As only one source (NGC 6240) is a ULIRG (L$_{IR}= 8.5 \times 10^{11}$), we hereafter refer to the sample as LIRGs. 
Figure \ref{fig:histograms} shows the integrated properties of the galaxies in this study (detailed in Table~\ref{tab:goals_sample}) in comparison with the entire GOALS sample. Our subsample is representative of the parent sample, though with limited sampling of the highest luminosity regime given that only one ULIRG is present. 

\subsection{\textit{Spitzer}/IRS observations}
\label{CUBISM}
Galaxies were mapped by carrying out spectral scans with the \textit{Spitzer}/IRS using low-resolution ($R{\sim}$60--120) slit spectroscopy in the SL (short-low) module (5.2--14.5 $\mu$m) and LL (long-low) module (14--38 $\mu$m). The data were retrieved from the \textit{Spitzer} Heritage Archive\footnote{\href{https://sha.ipac.caltech.edu/applications/Spitzer/SHA/}{https://sha.ipac.caltech.edu/applications/Spitzer/SHA/}} (PID 30323; PI L. Armus and archival data), which provides public access to raw, reprocessed, and calibrated observations. For this study, we collected all spectral mapping observations available to investigate the extended emission in our sample of LIRGs.
To build the spectral cubes, we processed the \textit{Spitzer}/IRS low-resolution spectral mapping observations using the CUbe Builder for IRS Spectral Mapping (CUBISM, \citealt{CUBISM}), which constructs spectral cubes from the grid of long-slit data. Based on the basic calibrated data (BCDs) from IRS maps, we generated 3D spectral cubes. Each module (SL and LL) contains two slits, resulting in an individual spectral cube for each slit. Consequently, for galaxies with mapping observations in all four low-resolution orders (SL2, SL1, LL2, and LL1), we generated four separate spectral cubes. The SL and LL modules differ in pixel scale, with the SL modules having a pixel scale of 1.85$\arcsec$/pixel with a point spread function (PSF) of $\sim$ 4$\arcsec$ and a field of view of 3.6$\arcsec$ $\times$ 57$\arcsec$, while the LL modules a pixel scale of 5.08$\arcsec$/pixel with a PSF of $\sim$ 11$\arcsec$ and a field of view of 10.5$\arcsec$ $\times$ 168$\arcsec$. We corrected the background emission in the 2D spectral images by using the secondary order of the slit as the background for the primary order, since each order is dispersed differently. Once the background is set for each order, the last step consists of manually removing the bad pixels. CUBISM automatically generates an additional uncertainty cube alongside the main spectral cube. 

\subsection{\textit{Herschel}/PACS observations}
\label{Herschel observations}
The Integral Field Spectrometer (IFS) of the PACS instrument on board \textit{Herschel} can perform simultaneous observations in the wavelength ranges of 51--73 $\mu$m and 70--105 $\mu$m (third and second orders, respectively; ``blue'' camera, $R{\sim}$1500--3000), and 102–210 $\mu$m (first order; ``red'' camera, $R{\sim}$1000--1500). The IFU features a $5 \times 5$ array of individual detectors (spaxels), each with a field of view (FOV) of $\sim 9.4\arcsec$ on a side, resulting in a total FOV of $47\arcsec \times 47\arcsec$. At the median distance of our sample, 84 Mpc, the PACS FOV corresponds to approximately 20 kpc on a side. Each pixel contains 16 spectral elements, which are combined using an image slicer across two $16 \times 25$ Ge:Ga detector arrays (blue and red cameras). The selected spectral range was scanned multiple times to achieve spectral resolution sufficient to meet or exceed Nyquist sampling. 

Most of the data were provided by OT1 and OT2 programs (OT1\_larmus\_1, OT2\_larmus\_1; P.I.: L.
Armus, accounting for more than 200 hr of observing time in
total), and from other programs available in the Herschel archive: KPGT\_esturm\_1 (P.I.: E. Sturm), KPOT\_pvanderw\_1 (P.I.:P. van der Werf), and OT1\_dweedman\_1 (P.I.: D. Weedman). Through PACS/IFS range spectroscopy, we obtained spectral map observations for the [O\,\textsc{i}]$_{63}$ and [C\,\textsc{ii}]$_{158}$ fine-structure emission lines available for 66 and 41 galaxies, respectively. For more detailed information, we refer the reader to \citet{Tanio13}. 

\color{black}
\section{Spectral extraction and emission line measurements}
\label{section3}
\subsection{Spectral extraction analysis}
\begin{figure*}[t] 
    \centering
    \includegraphics[width=0.85\textwidth]{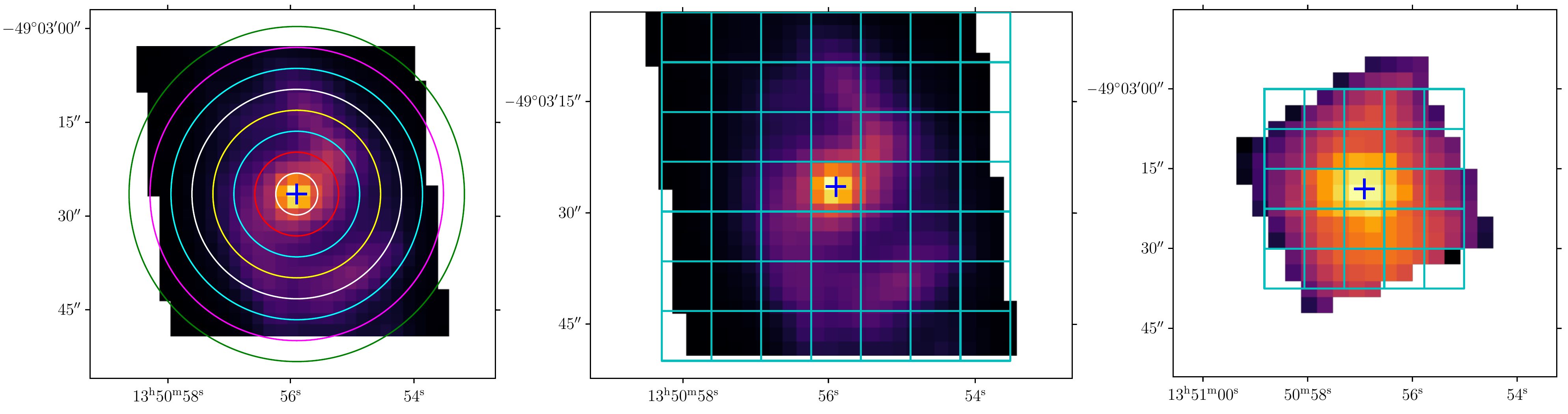}
    \caption{Example galaxy from the subsample, ESO221-IG010, illustrating the two different methodologies used for spectral extraction. Left: Concentric annuli around the centroid (blue cross) on \textit{Spitzer}/IRS observation. Each annulus has a physical width of 1 kpc. The largest aperture has a physical radius of 8 kpc. Middle: Square-aperture grid around the centroid on \textit{Spitzer}/IRS spectral map. The side of each square is 2 kpc (Spitzer grid). Both \textit{Spitzer}/IRS are shown at 13.3 $\mu$m. Right: Square-aperture grid around the centroid on CII \textit{Herschel}/PACS spectral map. The length of the aperture is 7.5$\arcsec$ (Herschel grid). All maps are reoriented to match the standard north-up, east-left orientation of the \textit{Herschel}/PACS maps.} 
    \label{fig:grids}
\end{figure*}

The aim of this analysis is to extract spectra from both the nuclear and extended regions of the galaxies. 
\label{spectral extraction} 
For this purpose, we used the \textit{Spitzer}/IRS cubes and a central circular aperture to extract the nuclear emission and concentric annuli for the extended emission. As discussed in Sect.~\ref{CUBISM}, each spectral order generates an individual cube covering a specific wavelength range. The centroid of each source was determined by fitting a 2D Gaussian to the source emission after collapsing the entire cube.
For merging systems, each galaxy is treated individually. The center is defined using the coordinates reported in (\citealt{Armus09}, see Table~\ref{tab:goals_sample}). Annular regions are then constructed around each member of the system. To avoid contamination between interacting galaxies, we visually inspect the cubes and exclude annuli that overlap with emission from neighboring sources.
For the most advanced mergers (stage 4), the individual components can no longer be reliably distinguished in the MIR emission. In these cases, the system is treated as a single source and a single center is defined. When selecting the radius of the annuli for a source, we take into account its angular distance, ensuring that the aperture encompasses the same physical area across all galaxies. Considering the luminosity distances spanned by our galaxy sample, we take a 2 kpc diameter aperture at $8\mu$m as a reference for the innermost (nuclear) extraction, while ensuring that the angular diameter of the aperture is always larger than the full width at half maximum of the PSF at wavelengths below $\sim$~32 $\mu$m for the most distant sources. The angular diameter of the aperture is described as $\Theta = 2\cdot r_{ap}$ and the physical diameter, D in Mpc, as $\Theta\cdot d_A$ where d$_A$ is the angular size distance. Given $\Theta$ in arcseconds, D=$\frac{d_L \cdot \Theta}{(1+z)^{2} \cdot 206265}$, where d$_L$ is the luminosity distance from \citet{Armus09} and $z$ is the redshift of the galaxy. We want the aperture to be D=2 kpc and we defined $\Theta = 2\cdot r_{ap}$, thus 
\begin{equation}
    r_{\rm ap}^{8\,\mu{\rm m}}(\arcsec) =
    \frac{10^{-3}\cdot 206265\cdot (1+z)^2}{d_{L}}
    \label{radius formula}
.\end{equation}
Following the computation of the centroid, we calculated the nuclear fluxes by integrating the intensity within the central aperture at each wavelength of the cubes. As regards to the flux uncertainty,  it is defined as $\sigma_{ap}=\sqrt{\sum_{i=1}^{N} {\sigma_i}^2 }$ where $\sigma_{i}$ is the uncorrelated flux uncertainty for each pixel in the region, provided by the uncertainty cubes from CUBISM (see Sect. \ref{CUBISM}). In this analysis, spectral data were first extracted from concentric annular regions, each with a width of 1~kpc, centered on each galaxy (see Sect.\ref{spectral extraction}). The extraction extends out to a physical distance of 8~kpc from the nucleus, corresponding to the maximum radial extent commonly covered across the sample, and is illustrated in Fig.~\ref{fig:grids}. To improve the signal-to-noise ratio (S/N), we subsequently aggregated these spectra into larger concentric regions. Specifically, we defined three distinct radial zones for this purpose: the nuclear region (from 0 to 2 kpc), the intermediate region (from 2 to 5 kpc), and the outer region (from 5 to 8 kpc) as shown in Figs.~\ref{fig:AGN_stacked}, \ref{fig:LIR_stacked}, \ref{fig:ssfr_stacked}, and~\ref{fig:merger_stacked}. All the spectra were median-stacked into property bins and arbitrarily normalized at 13.2 $\mu$m, in order to better highlight how the MIR spectra may be related to the integrated properties of galaxies. 

To capture variations in the ISM properties that are not radially symmetric and therefore may not change concentrically, we also defined a grid of square apertures across each galaxy to ensure complete spatial coverage. This approach allows us to separate and study specific regions of the galaxies, thereby enabling us to better trace the potentially local origin of the emission, which is not feasible with the averages obtained from the concentric ring analysis. Figure~\ref{fig:grids} shows that the physical side of the square regions is 2 kpc, with $r_{ap}^{8\mu m}(\arcsec)$ computed following Eq.~\ref{radius formula}. Henceforth, this specific grid will be referred to as the "Spitzer grid." We note that the World Coordinate System (WCS) of the galaxies observed with the \textit{Spitzer}/IRS has arbitrary orientations. To enable a direct comparison of the emission lines from the MIR and FIR, the images were rotated to have the standard north-up, east-left orientation, to match the \textit{Herschel}/PACS maps. To extract the flux from the \textit{Herschel}/PACS cubes, we adopted square apertures with a side length of 7.5$\arcsec$, slightly smaller than the spaxel size of 9.4$\arcsec$ (see Fig.~\ref{fig:grids}). 
We acknowledge that this introduces a small level of spatial correlation in the extracted line fluxes, but this is minimized by the fact that the Herschel PSF is subsampled, and is smaller than the \textit{PACS} spaxels, so the data are independent from spaxel to spaxel. This aperture size was chosen to maximize the spatial sampling of each galaxy given the limited FOV of \textit{Herschel}/PACS, as larger apertures would result in fewer measurements across the galactic extent.
In order to compare the MIR and FIR lines from the two different telescopes, we use this grid (square apertures with a side length of 7.5$\arcsec$) on the \textit{Spitzer}/IRS cubes. We refer to this configuration as the "Herschel grid." Therefore, we use the three aperture configurations for different purposes throughout the paper. The concentric annuli are used to construct the stacked radial \textit{Spitzer}/IRS spectra presented in Sect.~\ref{radial variation}. The square 2 kpc apertures, the 
Spitzer grid, are used for the spatially resolved MIR analysis illustrated in Sect.~\ref{PAH ratio} and the 7.5$\arcsec$ square apertures defined on the PACS maps and applied also to the IRS cubes, the Herschel grid, are used when comparing MIR and FIR measurements on matched spatial scales in Sect.~\ref{photo-effiency}. 

We note that for highly inclined systems, projection effects may lead to contributions from disk emission along the line of sight toward the nucleus, potentially diluting intrinsic radial trends. This represents a general limitation of the analysis. However, in our sample, most galaxies are either interacting systems without a well-defined disk geometry or are face-on (or close to face-on). For the subset of disk-like systems, only a small fraction (10 out of 67 galaxies) shows significant inclination, with 5 sources being highly inclined (close to edge-on) and 5 exhibiting moderate inclinations ($\sim$ 45$^{\circ}$). Therefore, projection effects are expected to have a limited impact on the overall radial trends presented in this work.

\subsection{Spectral fitting}
To model the MIR spectral features observed with \textit{Spitzer}/IRS, we used a modified version of the Continuum And Feature Extraction code (CAFE\footnote{\href{https://github.com/GOALS-survey/CAFE}{https://github.com/GOALS-survey/CAFE}}; \citealt{Marshall07, CAFE25}). The spectra are decomposed into PAH emission, atomic and molecular lines, molecular absorption, and dust continuum emission components. We present PAH and emission-line measurements corrected for dust attenuation, adopting a mixed geometry for the PAH component and a foreground screen geometry for the warm and hot dust continua, with configurations tailored to starburst- and AGN-dominated sources (see Appendix~\ref{cafe_fits}). 

Regarding the FIR spectra obtained from \textit{Herschel}/PACS, to accurately fit the emission lines, we used a Gaussian model to represent the line profiles and a linear continuum model to account for the continuum. Initially, the Gaussian and linear continuum model were fitted to the entire spectrum. The continuum component was then subtracted and a refined fitting was conducted within a $\pm~6\sigma$ region around the emission line wavelength, based on the initial fit. This two-step process facilitated a more precise determination of the emission line flux. Uncertainties are based on the spectrum of each line and for each aperture, capturing the uncertainties measured directly from the data. The line flux uncertainty was estimated using the covariance matrix from the Gaussian fit to the spectrum. The uncertainties of the model parameters were propagated to compute the flux uncertainty at each wavelength within the ${\pm}~3\sigma$ region around the line center. These individual uncertainties were then combined in quadrature to provide the total uncertainty in the integrated line flux. Absolute photometric errors can reach up to approximately 11\%--12\%\footnote{\href{ http://herschel.esac.esa.int/Docs/PACS/html/ch04s10.html}{ http://herschel.esac.esa.int/Docs/PACS/html/ch04s10.html}}, thus were taken into account and added in quadrature to the measured uncertainties from the individual spectra.

\section{Results}
\label{results}

\begin{figure*}[t] 
    \centering
    \includegraphics[width=0.9\textwidth]{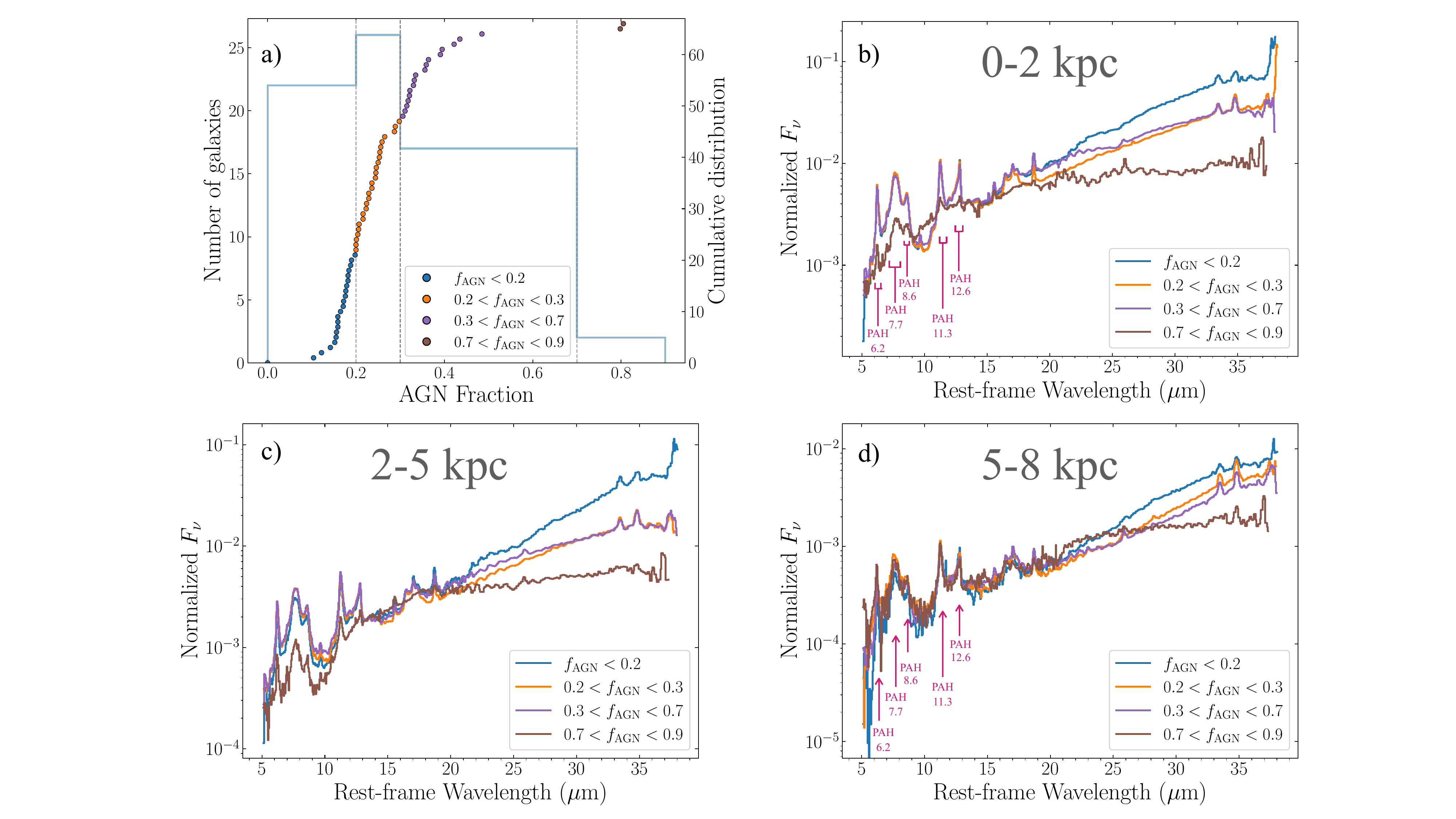}
    \caption{Galaxy sample divided into four bins according to their global f$_{\rm AGN}$. Panel (a): Cumulative distribution function of the galaxy sample that illustrates how galaxies are distributed across the four mid-IR f$_{AGN}$ bins. Panels (b), (c), and (d): AGN fraction-weighted spectra for each bin at different galactocentric distances: the nuclear region (0-2 kpc), the intermediate region (2-5 kpc), and the outer region (5-8 kpc), respectively. All spectra are shown in the rest-frame and normalized at 13.2 $\mu$m. Galaxies in the highest f$_{\rm AGN}$ bin show progressive diminishing AGN influence with increasing distance from the nuclear region.}
    \label{fig:AGN_stacked}
\end{figure*}

\subsection{Radial variation of stacked mid-infrared spectra}
\label{radial variation}
After extracting fluxes from the nuclear and extended regions of all galaxies in our sample using the concentric annuli method, we construct radial profiles of each gaseous ISM phase property by stacking the extracted spectra according to integrated galaxy properties. Our goal is to study how the ISM conditions change with the distance to their center of the galaxy as a function of the integrated properties of the galaxies and their main power source (IR luminosity, f$_{AGN}$, merger stage, and sSFR).

Figure~\ref{fig:AGN_stacked} shows the stacked spectra of the galaxies in our sample divided into four bins according to their global mid-IR f$_{AGN}$ listed in Table~\ref{tab:goals_sample}.  
As shown in panel (b) of Fig.~\ref{fig:AGN_stacked}, in the nuclear region (0-2 kpc), the hot dust continuum associated with the AGN significantly dilutes PAH features and emission lines and fills in the 9.7 $\mu$m silicate absorption feature. This phenomenon is well documented in the literature, particularly for strong AGNs with minimal obscuration, which produce nearly featureless spectra, predominantly attributed to hot dust emission \citep{Stern05, Alonso_Herrero06, Donley07, Li07, Tanio10, Mullaney11}. At intermediate distances (2--5~kpc; panel c of Fig.~\ref{fig:AGN_stacked}), the AGN-heated dust continuum becomes less pronounced, allowing PAH features to become moderately detectable. This trend continues in the outermost regions (5-8 kpc; panel d of Fig.~\ref{fig:AGN_stacked}), where stacked spectra display similar characteristics regardless of f$_{AGN}$, demonstrating that AGN effects diminish substantially at larger galactocentric 
distances. The same analysis was conducted for IR luminosity, sSFR, and merger stage with results presented in Appendix~\ref{Appendix1}. Further analysis of these results will be presented in future work.

In the following parts of this section, the analysis instead focuses on the results obtained from the square-aperture grids. As mentioned in Sect.~\ref{spectral extraction}, when we display plots including only MIR line ratios, we use the Spitzer grid configuration; when we present plots including both MIR and FIR line ratios, we adopt the Herschel grid configuration. 

\subsection{Spatially resolved PAH features} 
\label{PAH ratio}

\begin{figure*}[h]
        \centering
                \includegraphics[width=1\textwidth]{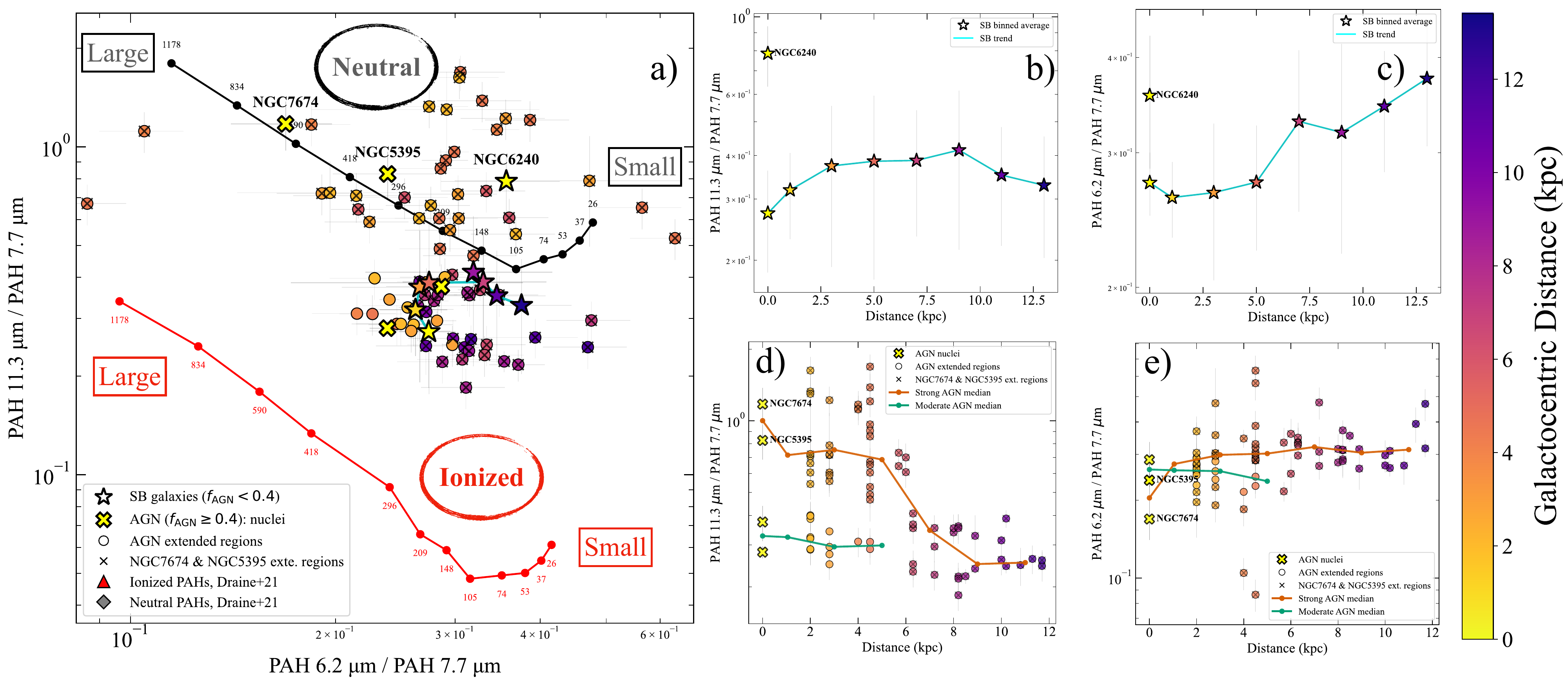}
                \caption{PAH diagnostic plot showing PAH~11.3~$\mu$m/7.7~$\mu$m versus 6.2~$\mu$m/7.7~$\mu$m ratios for nuclear and extended regions. The PAH~11.3/7.7 and 6.2/7.7 ratios primarily trace the PAH ionization state and characteristic grain size, respectively, with higher values indicating more neutral and smaller PAHs. Data points are color-coded according to their projected galactocentric distance. Star symbols denote starburst regions (f$_{\mathrm{AGN}}{<}0.4$), yellow crosses indicate AGN-dominated nuclei (f$_{\mathrm{AGN}}{\geq}0.4$), and circles represent the extended regions of AGN-dominated galaxies. Extended regions of the two most AGN-dominated sources, NGC~7674 and NGC~5395, are additionally marked with an overlaid cross ($\times)$ symbol. The error bars on the SB binned points represent the dispersion of the data within each radial bin. Theoretical PAH model grids from \citet{Draine21} are overlaid (black for fully neutral and red for fully ionized PAHs), computed for a dimensionless radiation field intensity of $\log(\rm U)$=2. For clarity and improved readability, we present the PAH~11.3/7.7 and 6.2/7.7 ratios separately for starburst- (Figs.~3b and 3c) and AGN-dominated systems (Figs.~3d and 3e). Within the inner $\sim$5\,kpc, both starburst- and AGN-dominated galaxies are characterized by PAH populations dominated by larger and more neutral grains. In starburst galaxies, the PAH size distribution shifts toward smaller grains beyond $\sim$5\,kpc, while a marked increase in PAH ionization is observed only at larger galactocentric distances ($\gtrsim$8\,kpc). In contrast, AGN-dominated systems exhibit a more rapid transition (at $\sim$5~kpc) toward smaller and increasingly ionized PAHs with radius. Overall, these trends indicate that the dominant power source, whether star formation or AGN activity, regulates PAH emission properties on kiloparsec scales.}
        \label{fig:pah_ratio}
\end{figure*}

Figure~\ref{fig:pah_ratio} shows the distribution of PAH band ratios for our galaxy sample that can be used as diagnostics of both the PAH ionization state and average grain size. The 7.7~$\mu$m complex predominantly traces ionized PAHs, while the 11.3~$\mu$m band is primarily associated with neutral PAH grains \citep{Draine07,Smith07}. In addition, these bands are sensitive to PAH size, with longer-wavelength features generally arising from larger molecules \citep{Galliano08}. 

Individual regions are distinguished by symbol shape and color. The projected radial distance from the galaxy center is indicated by color. Starburst-dominated galaxies (f$_{\mathrm{AGN}}{<}0.4$) are marked with star symbols for both nuclear and extended regions. In this case, each star corresponds to the mean PAH ratios computed from all starburst regions located at a given radial distance, and the associated error bars represent the dispersion of the data within each bin, reflecting the intrinsic scatter of the distribution. To aid the reader in visualizing the radial behavior of PAH properties, we present the radial trends of the PAH~11.3/7.7 and PAH~6.2/7.7 ratios for both starburst galaxies (Figs.~\ref{fig:pah_ratio}b and \ref{fig:pah_ratio}c) and AGN-host systems (Fig.~\ref{fig:pah_ratio}d and \ref{fig:pah_ratio}e).   
AGN-dominated systems (f$_{\mathrm{AGN}}{\geq}0.4$) are distinguished by different symbols for nuclear and extended regions. AGN nuclei are shown as yellow crosses, while their extended regions are shown as circles. Among the AGN-dominated systems, NGC~7674 and NGC~5395 exhibit the highest AGN fractions, with f$_{\mathrm{AGN}}$=0.81 and 0.79, respectively, whereas NGC~5990 and NGC~7469 show more moderate AGN contributions of f$_{\mathrm{AGN}}$=0.49 and 0.44 (see Table~\ref{tab:goals_sample}). To emphasize this distinction, the extended regions of NGC~7674 and NGC~5395 are additionally marked with an overlaid "$\times$" symbol. The figure also includes theoretical PAH model tracks from \citet{Draine21}, which illustrate the expected evolution of PAH ratios as a function of ionization state and grain size. These tracks are based on models computed for a dimensionless radiation field intensity of $\log(\rm U)$=2, representative of moderately intense heating conditions (see Table~\ref{tab:pah_ratios_NC}). The reference model spectra used for comparison are taken from \citet{Draine21b}. All spectra were fitted using \textsc{CAFE}, ensuring a consistent and direct comparison between the models and our measurements. In Appendix~\ref{PAH from Draine21} we tabulate the PAH band ratios computed from \citet{Draine21} theoretical spectra for both neutral and ionized PAHs, spanning three grain size distributions (small, standard, large), three ionization functions (low, standard, high), and radiation field intensities logU=0--6 using CAFE.\\ [5pt]
\indent Starburst galaxies: 
From the nuclear regions out to $\sim$5~kpc, the extended regions of starburst-dominated galaxies show a systematic radial trend in PAH ionization (Fig.~\ref{fig:pah_ratio}, Fig.~\ref{fig:pah_ratio}b). In this inner regime, the PAH~11.3/7.7 ratio slightly increases with galactocentric distance (PAH~11.3/7.7 from $\sim$0.25 up to $\sim$0.4), indicating a progressively larger contribution from neutral PAHs. Over this radial range, the PAH~6.2/7.7 ratio remains approximately constant, indicating no significant variation in PAH size (PAH~6.2/7.7 $\sim$0.27, Fig.~\ref{fig:pah_ratio}c). Between $\sim$5 and $\sim$8~kpc, the PAH~11.3/7.7 ratio, Fig.~\ref{fig:pah_ratio}b, continues to increase, reaching values of $\sim$0.45, while the PAH~6.2/7.7 ratio, Fig.~\ref{fig:pah_ratio}c, begins to rise, signaling the onset of a transition toward smaller PAH sizes. Beyond $\sim$8~kpc, the radial trend in PAH ionization reverses. At larger galactocentric distances (8--13~kpc), the PAH~11.3/7.7 (Fig.~\ref{fig:pah_ratio}b) ratio decreases with increasing radius (from $\sim$0.45 down to $\sim$0.3), indicating increasingly ionized PAH populations. In this outer regime, PAHs are systematically smaller than in the inner regions, with PAH~6.2/7.7 varying from $\sim$0.33 to $\sim$0.4 (Fig.~\ref{fig:pah_ratio}c). This behavior is consistent with a photo-destruction scenario in which the intense ultraviolet radiation fields, produced by young, massive stars \citep{Smith07,Pereira-Santaella10} in the vicinity of the central starburst preferentially destroy smaller PAHs, while only larger PAH molecular chains are able to survive \citep{Voit92, Zhen15, Lange25}.
\\ [5pt]
\indent Active galactic nucleus-dominated galaxies:
AGN-dominated nuclei (yellow crosses) are broadly distributed across the PAH ratio diagram. The nuclear regions of the two galaxies with moderate AGN fractions, NGC~5990 and NGC~7469, overlap with the locus of nuclear starburst regions and are characterized by relatively low PAH~11.3/7.7 ratios (${\sim}$0.27 and ${\sim}$0.4, respectively, Fig.~\ref{fig:pah_ratio}d), in agreement with the resolved study of \citet{Donnelly24}, where they found an average PAH~11.3/7.7 ratio of ${\sim}$0.5 near the nucleus and in the bulge of a low-luminosity AGN. In comparison, the nuclei of the two galaxies hosting the most powerful AGNs, NGC~5395 and NGC~7674, are significantly offset toward higher PAH~11.3/7.7 values (${\sim}$0.8 and ${\sim}$1, respectively), indicating PAH populations that are more neutral and systematically larger (PAH~6.2/7.7 ratios ${\sim}$0.25 and ${\sim}$0.17, respectively, Fig.~\ref{fig:pah_ratio}e) than those associated with the lower-f$_{\mathrm{AGN}}$ systems. This pronounced separation suggests that clear modifications of the nuclear PAH emission properties occur only in galaxies hosting strong AGN activity, consistent with scenarios in which the harsh AGN radiation field selectively destroys ionized PAHs thus favoring the survival of neutral \citep{Smith07,Sabrina14,Garcia-Bernete22} and, larger \citep{McKinney21,Lai22,Lai23} grains. The extended ISM of the most powerful AGNs (marked by an overlaid "$\times$" symbol) further supports this interpretation. Regions located at intermediate radii (1–5~kpc) reach PAH~11.3/7.7 ratios (Fig.~\ref{fig:pah_ratio}d) comparable to those observed in the nuclei (0.4${\leq}$PAH~11.3/7.7${\leq}$~1.5), implying that the AGN influence extends well beyond the central kiloparsec and significantly alters PAH populations on several kiloparsec scales. At larger distances (6--13~kpc), the extended regions of even the most AGN-dominated overlap or show lower PAH~11.3/7.7 ratios than the nuclear star-forming locus (0.17${\leq}$PAH~11.3/7.7${\leq}$~0.4), indicating a progressive weakening of the AGN impact with radius.
Interestingly, the extended regions of moderate-AGN galaxies at 1–6~kpc occupy a parameter space similar to that of the outer (6–14~kpc) regions of the most powerful AGNs. This suggests that strong AGNs can affect PAH ionization states out to larger galactocentric distances, with the same effectiveness that moderate AGNs achieve at smaller scales (1--6~kpc). Remarkably, extended regions of moderately AGN-dominated systems lie closer to the starburst locus yet remain distinct and do not overlap with it. This is consistent with the results of \citet{Donnelly24}, who found that the impact of a low-luminosity AGN on PAH ionization and size might be confined to the innermost regions, with no significant effect at larger kiloparsec scales. Expanding the sample of both strong and moderate AGNs, together with higher-spatial-resolution observations (e.g., with JWST), would help to further clarify this behavior. It is worth also mentioning that NGC~6240 occupies a distinct region of the diagram compared to the other starburst-dominated galaxies. This is likely related to its ULIRG nature and the presence of a heavily obscured AGN, which results in more extreme ISM conditions. As the only ULIRG in our sample, it is not representative of the general trends discussed here, and a detailed analysis of its properties is beyond the scope of this work. 

The dependence of PAH emission and PAH band ratios on gas-phase metallicity has been widely documented in previous studies, and could in principle contribute to the radial trends observed in our analysis \citep[e.g.,][]{Smith07, Shivaei24, Whitcomb24}. However, in merger and interacting systems, tidal forces and gas inflows efficiently redistribute the gas within galaxies, leading to significantly flattened metallicity gradients \citep[e.g.,][]{Rupke10, Kewley10, Rich12} compared to those observed in isolated spiral galaxies.
We have independently investigated the radial behavior of the gas-phase metallicity for the same galaxies in the companion paper (Paper II), where we combine MIR and FIR fine-structure line diagnostics and interpret them using \texttt{CLOUDY} photoionization models. That analysis shows that both starburst- and AGN-dominated systems exhibit approximately flat metallicity profiles, with typical values Z/Z$_{\odot}$ $\sim$ 1.5–1.7 and no statistically significant radial gradients over the spatial scales probed.
These results indicate that metallicity variations are unlikely to be the primary driver of the observed radial trends in PAH band ratios. Instead, the variations are more naturally explained by changes in the radiation field properties and local ISM conditions, as discussed above. 

Overall, both starburst- and strong AGN-dominated galaxies exhibit a common radial behavior in their PAH properties. In the inner regions, out to $\sim$5~kpc, the PAH population is characterized by larger and more neutral grains, whereas at larger galactocentric distances the PAHs become progressively more ionized and smaller. This indicates that the dominant power source, whether associated with star formation or AGN activity, can significantly influence the PAH emission properties of the extended ISM out to scales of $\sim$5~kpc; although powerful AGNs have a significantly higher impact than star formation, increasing the 11.3/7.7 PAH ratio by factors of a few with respect to the overall location of star-forming regions and reaching larger distances. In this case as well, a photo-destruction scenario remains plausible in the presence of the intense radiation field associated with the AGN \citep{Diamond10, Garcia-Bernete22, Lai22}.

\begin{figure*}[ht]
    \centering
    \begin{subfigure}[t]{0.49\textwidth}
        \centering
        \includegraphics[width=\linewidth]{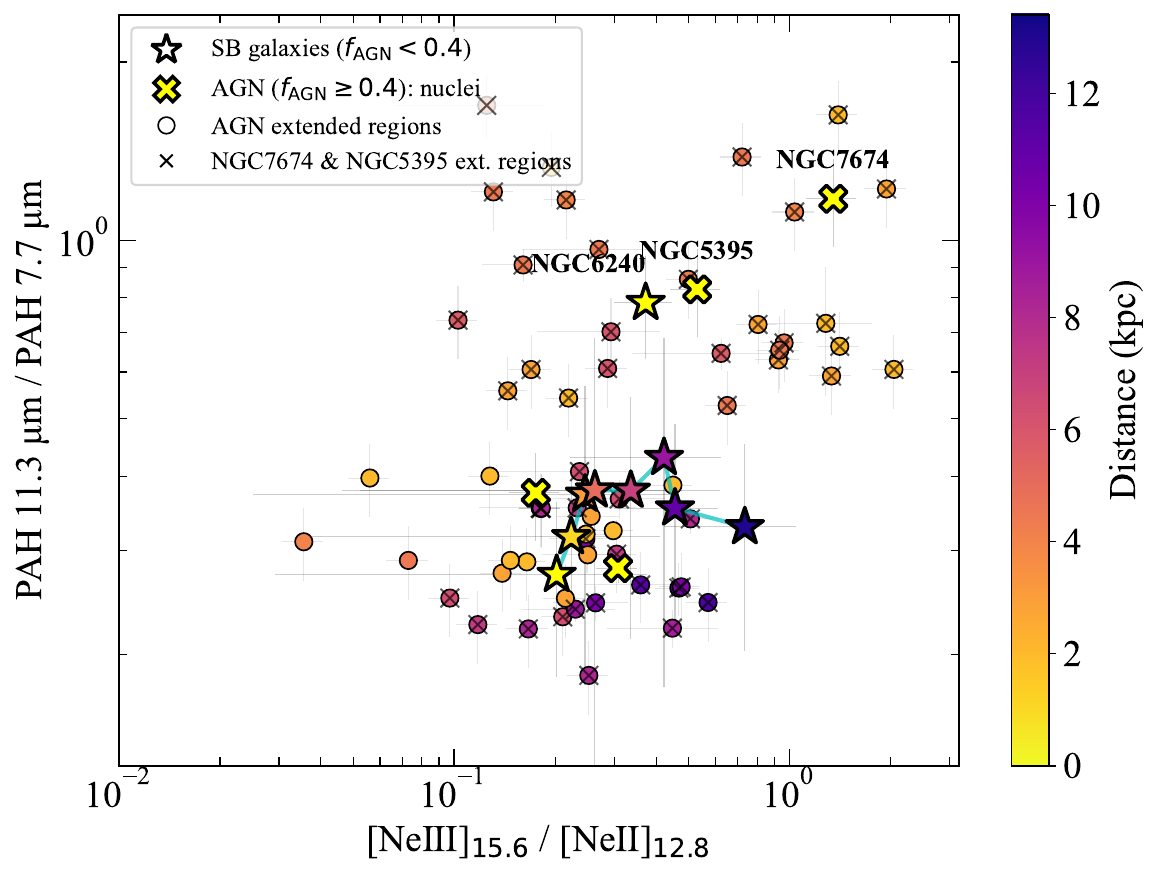}
        \label{fig:PAH11-7_Ne}
    \end{subfigure}
    \hfill
    \begin{subfigure}[t]{0.49\textwidth}
        \centering
        \includegraphics[width=\linewidth]{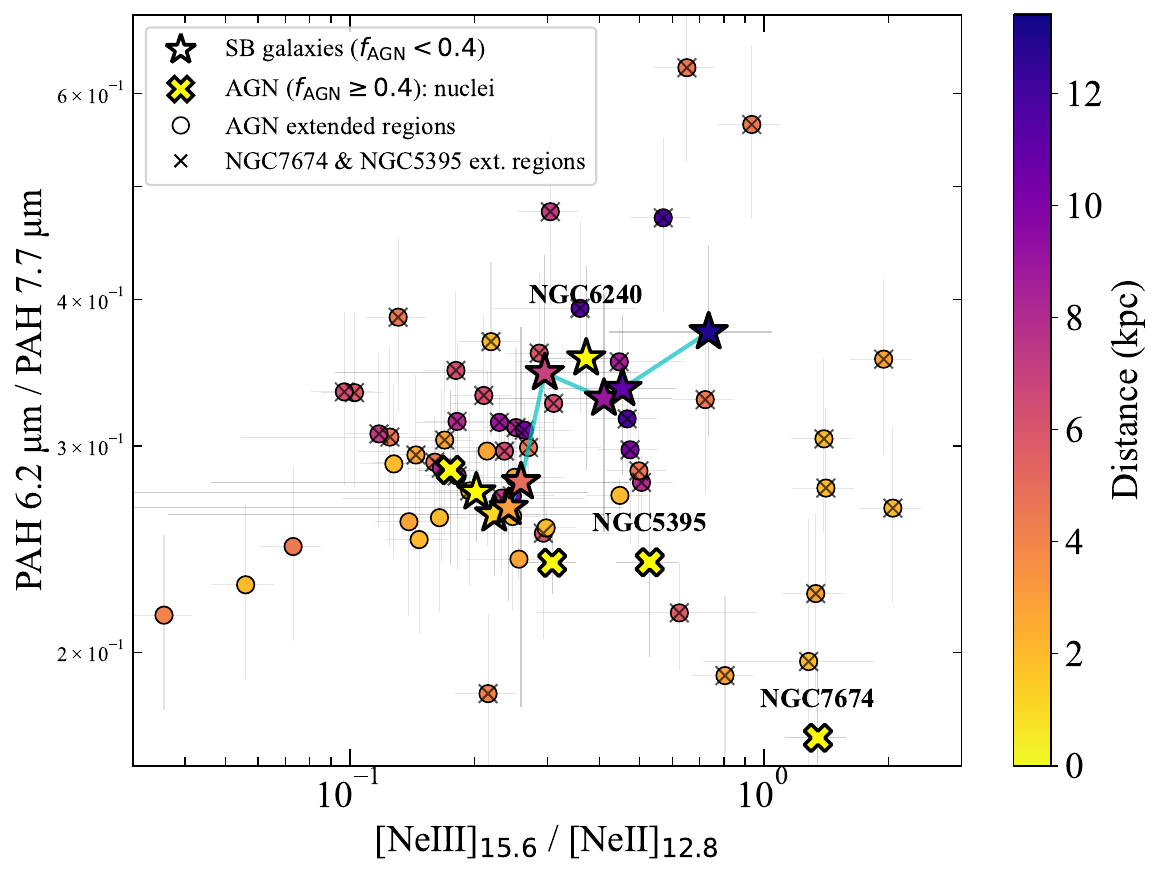}
        \label{fig:PAH62-7_Ne}
    \end{subfigure}
    \vspace{-0.3cm}
    \caption{Distribution of PAH ratios as a function of the intensity of the radiation field traced by [NeIII]$_{15}$/[NeII]$_{13}$. Left: PAH~11.3~$\mu$m/7.7~$\mu$m. Right: PAH~6.2~$\mu$m/7.7~$\mu$m. Symbols follow the same convention as in Fig. \ref{fig:pah_ratio}. Starburst nuclei show low PAH~11.3/7.7 ratios (ionized PAHs) and constant PAH~6.2/7.7 ratios despite varying [NeIII]$_{15}$/[NeII]$_{13}$, transitioning to smaller grains at larger distances. AGNs nuclear regions exhibit elevated PAH~11.3/7.7 ratios (neutral PAHs) and extended influence to kiloparsec scales, with NGC~7674 showing PAH destruction signatures.}
    \label{fig:pah_vs_neon}
\end{figure*}

\subsection{The effect of the radiation field}
The left panel of Fig.~\ref{fig:pah_vs_neon} explores the relationship between the PAH ionization state, traced by the PAH~11.3/7.7 ratio, and the hardness of the ionizing radiation field \citep[e.g.,][]{Madden06, Decarli-diaz25}, as probed by the [NeIII]$_{15}$/[NeII]$_{13}$ line ratio. To guide the reader, we also plot the radial trend of the ratio in Fig.~\ref{fig:NeIII_NeII_vs_dist} for starburst- (left panel) and for AGN-dominated galaxies (right panel). These lines originate from ions with ionization potentials of 41.0~eV and 21.6~eV, respectively. The right panel of Fig.~\ref{fig:pah_vs_neon} examines the dependence of the PAH~6.2/7.7 ratio on the same line ratio of radiation-field intensity. \\ [5pt]
\indent Starburst galaxies.
Starburst-dominated systems exhibit a radial variation in their PAH ionization properties. From the nuclear regions out to $\sim$8~kpc, the PAH~11.3/7.7 ratio slightly increases with distance, indicating a growing contribution from neutral PAHs (Sect.~\ref{PAH ratio} and see also \citealt{Pereira-Santaella10}). At larger radii ($>$8~kpc), the radiation field intensity increases (from [NeIII]$_{15}$/[NeII]$_{13}$${\sim}$0.3 to ${\sim}$0.9), whereas no statistically significant variation is observed in the PAH ratio (left panel).
In contrast, the PAH~6.2/7.7 ratio exhibits a clear change beyond $\sim$5~kpc (right panel), with these regions showing a higher relative contribution from smaller PAHs (PAH~6.2/7.7 from ${\sim}$0.27 to ${\sim}$0.4). This behavior coincides with an increase in the [Ne\,\textsc{iii}]$_{15}$/[Ne\,\textsc{ii}]$_{13}$ ratio, suggesting that variations in the radiation field intensity can influence the PAH size distribution.
This behavior may result from a combination of enhanced ionizing radiation and fast AGN-driven winds, which can preferentially destroy smaller PAH grains \citep{Lai22}.

Overall, variations in the PAH ionization state are modest, with the binned values spanning a relatively narrow range and appearing only weakly associated with changes in the radiation field intensity. 
Instead, the PAH size distribution shows a stronger correlation with the radiation field strength, supporting a scenario in which smaller PAHs are preferentially destroyed while larger species remain intact.
 \\ [5pt]
\indent Active galactic nucleus-dominated galaxies.
In contrast to the starburst systems, AGN-dominated nuclei exhibit a broader distribution in both PAH ratios and radiation field intensity. Nuclei hosting moderate AGN fractions overlap with the starburst locus, whereas the most AGN-dominated sources, NGC~5395 and NGC~7674, are clearly offset toward higher [NeIII]$_{15}$/[NeII]$_{13}$ ratios ($\sim$0.5 and $\sim$1.5, respectively). These nuclei are characterized by elevated PAH~11.3/7.7 ratios together with suppressed PAH~6.2/7.7 ratios, a behavior that is particularly pronounced in NGC~7674. In this source, the markedly low PAH~6.2/7.7 ratio ($\sim$ 0.17) suggests a suppression of the 6.2~$\mu$m feature, potentially driven by PAH dehydrogenation or destruction processes in the intense AGN radiation field \citep{Maragkoudakis18,Garcia-Bernete22, Lai22}. Overall, this combination of high PAH~11.3/7.7 and low PAH~6.2/7.7 ratios is consistent with selective processing of small, ionized PAHs and the relative survival of larger, more neutral grains \citep{Smith07,Garcia-Bernete22} in the presence of a hard radiation field. This interpretation is also consistent with spatially resolved studies showing that variations in PAH band ratios are strongly influenced by changes in the hardness of the radiation field exciting PAH grains \citep{Donnelly24}.
In moderate AGN systems, the extended regions display a slight decline in the radiation field intensity ([Ne\,\textsc{iii}]$_{15}$/[Ne\,\textsc{ii}]$_{13}$ decreasing from $\sim$0.2 to $\sim$0.04),  while no significant variations are observed in either the PAH ionization state or grain size within the radial range probed ($\lesssim$ 5 kpc). 
In contrast, the extended regions of the most AGN-dominated galaxies span a much broader range of [Ne\,\textsc{iii}]$_{15}$/[Ne\,\textsc{ii}]$_{13}$ values. At intermediate radii (1--5~kpc), these regions are characterized by high PAH~11.3/7.7 ratios (see Sect.~\ref{PAH ratio}) together with elevated [Ne\,\textsc{iii}]$_{15}$/[Ne\,\textsc{ii}]$_{13}$ values, comparable to those measured in the nuclear regions, whereas over the same radial range the PAH~6.2/7.7 ratio increases significantly. Beyond $\sim$6~kpc, the [Ne\,\textsc{iii}]$_{15}$/[Ne\,\textsc{ii}]$_{13}$ ratio converges toward values typical of starburst regions, accompanied by lower PAH~11.3/7.7 ratios and an approximately constant PAH~6.2/7.7 ratio relative to the inner regions. We note that strongly AGN-dominated systems extend to larger galactocentric distances compared to the moderate ones; however, this does not affect the comparison presented here, which is restricted to the shared radial range. We find that the [Ne\,\textsc{iii}]$_{15}$/[Ne\,\textsc{ii}]$_{13}$ ratio decreases with galactocentric distance for both strong and moderate AGNs. This behavior reflects the progressive weakening of the AGN radiation field with distance from the nucleus. As the AGN contribution declines, the radiation field becomes increasingly dominated by stellar emission, which is softer (e.g., \citealt{Pereira-Santaella10, Lai22}). This interpretation is supported by the radial profiles of the ionization parameter derived in Paper II, which show a systematic decrease with galactocentric distance in AGN-hosting systems.
Taken together, these trends indicate that strong AGN activity can harden the radiation field out to $\sim$5~kpc, primarily affecting the PAH size distribution without significantly altering the ionization state. At larger radii, the radiation field becomes increasingly attenuated, leading to a more ionized PAH population while the characteristic grain size remains largely unchanged. 

\section{Photoelectric efficiency}
\label{photo-effiency}
The photoelectric effect on dust grains is widely recognized as one of the dominant gas heating mechanisms in the neutral ISM, particularly in PDRs (e.g., \citealt{Bakes94, Hollenbach99}). In these environments, photoelectric heating, driven primarily by PAHs, regulates the thermal balance of the neutral ISM \citep{Bakes94, Weingartner01}, while cooling is dominated by the [C\,\textsc{ii}]$_{158}$ and [O\,\textsc{i}]$_{63}$ fine-structure lines \citep{Hollenbach99, Malhotra01, Croxall12, Beirao12, Rosenberg15, Tanio17, McKinney21}.

The photoelectric efficiency, $\epsilon_{\rm PE}$, is defined as the fraction of the absorbed far-ultraviolet (FUV) energy that is converted into gas heating via photoelectrons, relative to the total energy absorbed by dust grains \citep{Bakes94}. This efficiency depends sensitively on grain size, composition, and charge state, as well as on the local radiation field strength, and gas density
\citep[][]{Bakes94, Weingartner01}. Observationally, the photoelectric efficiency cannot be measured directly, but is commonly traced using the ratio of gas cooling to dust heating \citep{Malhotra01, Croxall12}. Since PAHs efficiently absorb FUV photons and re-radiate this energy in the mid-infrared \citep{Tielens08}, while the gas cools predominantly via far-IR fine-structure lines, a widely adopted proxy for $\epsilon_{\rm PE}$ is given by
\begin{equation}
\epsilon_{\rm PE}
\;\propto\;
\frac{
L_{\rm [C\,II],PDR} + L_{\rm [O\,I]}
}{
L_{\rm PAH}
},
\label{eq:epsilon_pe_def}
\end{equation}
where $L_{\rm [C\,II],PDR}$ is the [C\,II] luminosity arising from neutral PDR gas, $L_{\rm [O\,I]}$ is the [O\,I]$_{63}$ luminosity, and $L_{\rm PAH}$ is
the total PAH emission \citep{Malhotra01, Beirao12, Croxall12}. In this work, $L_{\rm PAH}$ is defined as the sum of the luminosities of the PAH~6.2~$\mu$m, PAH~7.7~$\mu$m, and PAH~11.3~$\mu$m emission features, which
together account for the bulk of the PAH power and dominate the MIR PAH emission in star-forming galaxies
\citep{Smith07, Tielens08}.
Because the [C\,II]$_{158}$ line can originate from both neutral and ionized gas, it is necessary to isolate the PDR-associated component before using it as a tracer of photoelectric heating \citep{Croxall17, Tanio17}. Following the approach of \citet{Tanio17} and \citet{McKinney21}, we computed the PDR contribution to the [C\,II] emission as
\begin{equation}
L_{\rm [C\,II],PDR}
=
\rm f_{\rm PDR} \times L_{\rm [C\,II],total},
\end{equation}
where f$_{\rm PDR}$ is the fraction of the total [C\,II] emission arising
from neutral PDRs. The PDR fraction f$_{\rm PDR}$ is estimated using the FIR color $S_{63}/S_{158}$ computed on the Herschel grid and following the empirical calibration derived by \citet{Tanio17}, which links the FIR continuum color to the relative contribution of neutral gas phase to the observed [C\,II] emission. \\

\begin{figure*}[ht]
    \centering
    \begin{subfigure}[t]{0.49\textwidth}
        \centering
        \includegraphics[width=\linewidth]{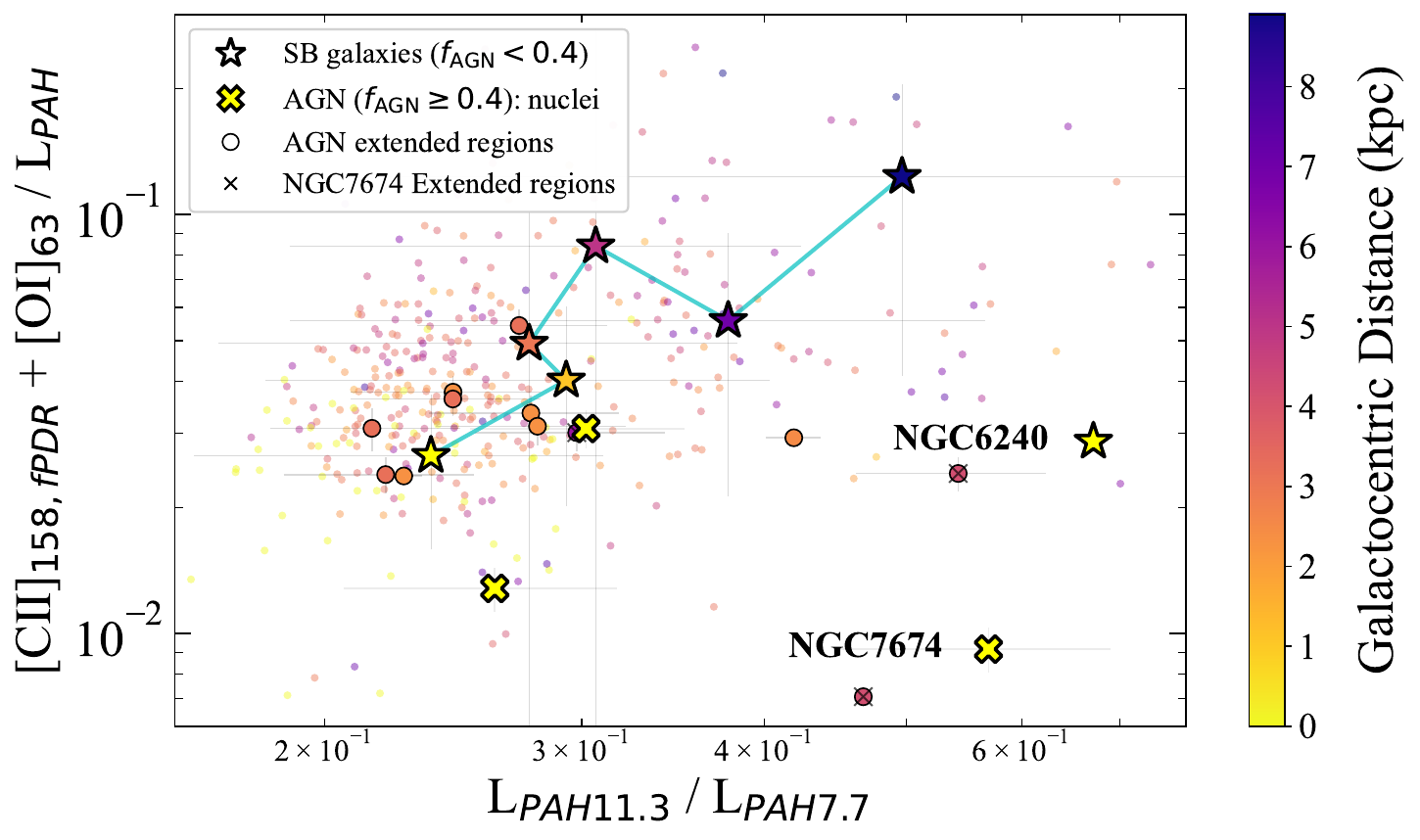}
        \label{fig:PAH11-7_Ne}
    \end{subfigure}
    \hfill
    \begin{subfigure}[t]{0.49\textwidth}
        \centering
        \includegraphics[width=\linewidth]{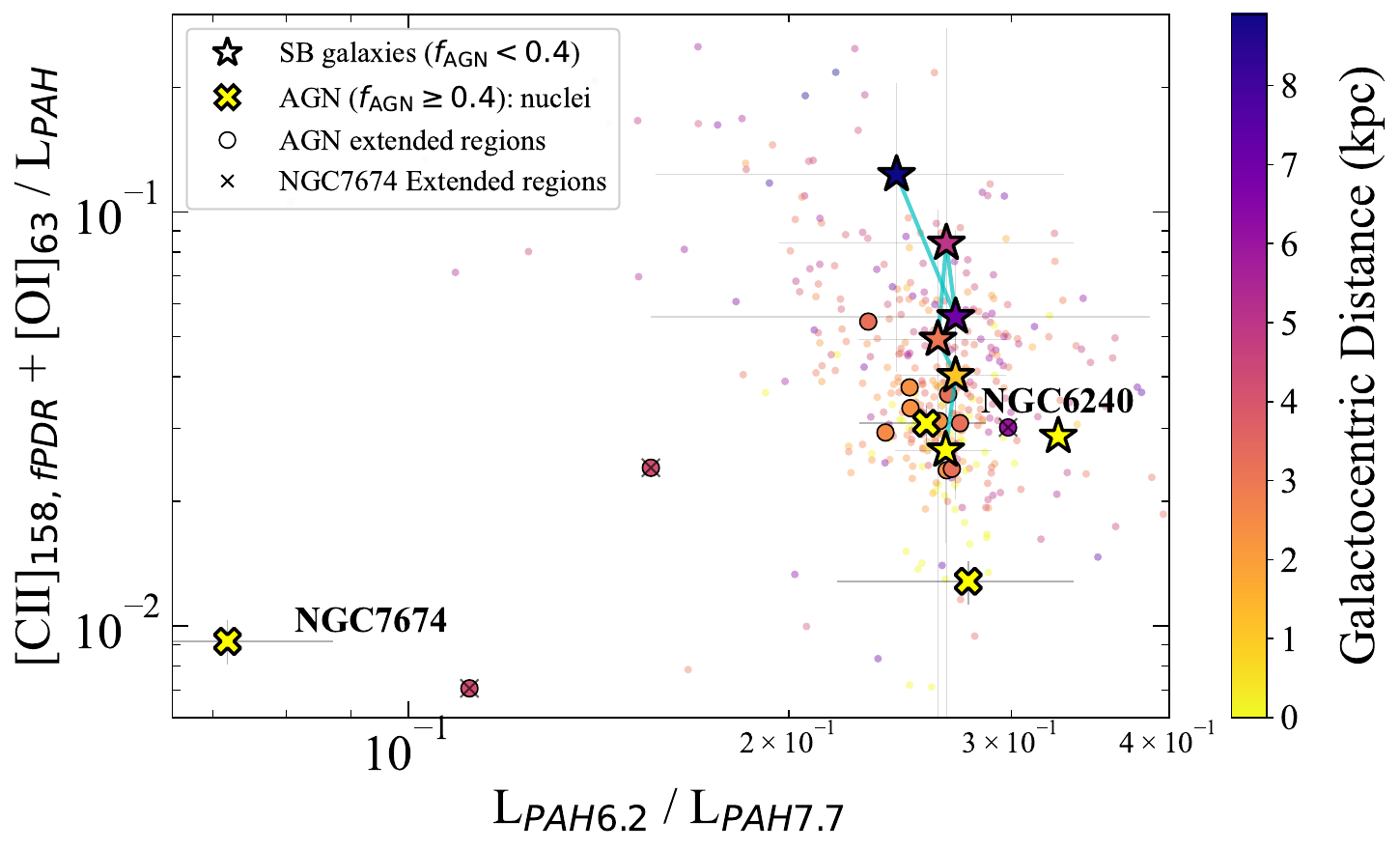}
        \label{fig:photo_vs_PAH}
    \end{subfigure}
    \vspace{-0.3cm}
    \caption{Photoelectric-efficiency proxy, $([{\rm C\,II}]_{\rm PDR}+[{\rm O\,I}])/L_{\rm PAH}$, as a function of PAH~11.3/7.7 (left) and PAH~6.2/7.7 (right). Symbols follow the same convention as in Fig. \ref{fig:pah_ratio}. The background points represent the full sample, color-coded according to their galactocentric distance. Overall, both starburst- and AGN-hosting galaxies show an increase in photoelectric efficiency with galactocentric distance, which indicates that central, intense radiation fields suppress efficient gas heating, while conditions in the outer regions favor higher photoelectric efficiency.}
    \label{fig:photo_vs_PAH}
\end{figure*}

Variations in the photoelectric efficiency are primarily driven by changes in the local radiation field and by the properties of the dust grains that dominate the photoelectric heating, in particular PAHs and very small grains
\citep{Bakes94, Weingartner01, Tielens08}. In many environments, increasing FUV radiation fields tend to shift the PAH charge balance toward more positively charged states, which can reduce the photoelectric yield and the efficiency with which absorbed UV energy is transferred to the gas \citep{Bakes94, Malhotra01}. However, the PAH charge distribution is not determined by the radiation field alone, but also depends on local gas properties such as the electron density, and temperature, as illustrated in regions such as Orion \citep[e.g.,][]{Maragkoudakis26}. Consequently, even in
regions where dust heating remains strong, the efficiency of gas heating can be suppressed, leading to lower values of $\epsilon_{\rm PE}$ \citep{Croxall12, McKinney21}.
In addition to charge, the PAH size distribution can influence
$\epsilon_{\rm PE}$, as smaller PAHs generally contribute more efficiently to the gas per absorbed photon than larger grains \citep{Bakes94,Weingartner01}. Changes in the local radiation field and ISM
conditions may therefore affect $\epsilon_{\rm PE}$ both by modifying the charge state of PAHs and by altering the relative contribution of different PAH sizes to the heating process \citep{Tielens08}. 

Figure~\ref{fig:photo_vs_PAH} shows the photoelectric efficiency, $\epsilon_{\rm PE}$, as a function of the PAH~11.3/7.7 ratio (left panel) and the PAH~6.2/7.7 ratio (right panel), tracing the PAH ionization state and grain size, respectively. The star symbols indicate the mean values within each galactocentric distance bin, the error bars indicate the dispersion of the data within each radial bin, while the background points display all individual measurements, color-coded by galactocentric distance. \\ [5pt]
\indent Starburst galaxies:
A positive correlation is observed between the photoelectric-efficiency proxy and the PAH~11.3/7.7 ratio. The photoelectric efficiency increases with galactocentric distance, accompanied by a rise in PAH~11.3/7.7. Beyond $\sim$5~kpc, the median relation becomes steeper, although the scatter increases at larger radii. This radial increase in photoelectric efficiency is consistent with spatially resolved studies showing that the efficiency varies across galactic environments and is higher in more diffuse regions than in intense star-forming regions (e.g., \citealt{Rubin09}). In our data, this trend is closely linked to the decrease of PAH ionization, probably due to the weakening of the average interstellar radiation field with distance (see Paper~II). From the nuclear regions toward larger galactocentric distances, the radiation field gradually weakens, leading to progressively more neutral PAH populations and a corresponding increase in photoelectric efficiency. This interpretation is supported by the radial profile of the ionization parameter presented in Paper~II, showing that starburst galaxies exhibit a declining ionization parameter with increasing galactocentric distance. This connection between PAH ionization and photoelectric efficiency is supported previous studies showing that grain charging plays a key role in regulating the efficiency of the photoelectric effect, with positively charged grains being less efficient at transferring energy to the gas (e.g., \citealt{Croxall12}). 

In Fig.~\ref{fig:photo_vs_PAH} (right panel), in contrast to the behavior observed for the PAH~11.3/7.7, the PAH~6.2/7.7 ratio remains approximately constant, while the photoelectric efficiency continues to rise. This suggests that the radial increase in photoelectric efficiency is not driven by changes in the PAH size distribution. Instead, the efficiency increases despite a constant PAH~6.2/7.7 ratio, reinforcing the idea that the dominant factor regulating photoelectric efficiency in these regions is the decreasing intensity of the radiation field with distance from the center and therefore the ionization state of the PAHs, rather than variations in PAH grain size. \\ [5pt] 
\indent Active galactic nucleus-dominated galaxies:
In the left panel of Fig.~\ref{fig:photo_vs_PAH}, AGN-hosting galaxies generally exhibit lower photoelectric efficiencies in their nuclear regions, followed by an increase with galactocentric distance. This radial trend is in line with spatially resolved studies showing that photoelectric efficiency increases away from regions characterized by more intense radiation fields (e.g., \citealt{Rubin09}; \citealt{Croxall12}). However, the behavior of AGN-hosting systems is more complex. In particular, the nucleus of NGC~7674 occupies the low $\epsilon_{\rm PE}$, high PAH~11.3/7.7 regime, pointing to relatively neutral PAHs in the central region. Moving to larger galactocentric distances, the photoelectric efficiency increases, while the PAH~11.3/7.7 ratio decreases, suggesting that PAHs become progressively more ionized with distance from the nucleus. This behavior contrasts with that observed in starburst galaxies, where higher photoelectric efficiency is associated with more neutral PAHs. However, as our analysis of the outer region of NGC 7674 relies on a single data point, these findings should be considered tentative. AGN-dominated galaxies with more moderate AGN fractions do not exhibit a strong radial variation in PAH ionization state, as traced by PAH~11.3/7.7, although they still show a systematic increase in photoelectric efficiency with galactocentric distance. These results indicate that, in AGN-hosting systems, the radial increase in photoelectric efficiency is primarily governed by the decline of the
radiation-field intensity with distance from the nucleus, as observed in starburst galaxies. This interpretation is also in agreement with previous resolved studies showing that some AGN-hosting systems exhibit a central suppression of photoelectric-efficiency proxies, although this effect is not uniquely linked to AGN activity and may instead reflect local ISM conditions such as radiation field intensity (e.g., \citealt{Smith17}). Nevertheless, the limited sample size for AGN-dominated systems, particularly in the photoelectric heating efficiency analysis, where the requirement of combined \textit{Spitzer} and \textit{Herschel} observations reduces both the number of sources and the spatial coverage, requires a cautious interpretation when linking photoelectric efficiency to PAH ionization state. 

In the right panel of Fig.~\ref{fig:photo_vs_PAH}, moderate AGN-hosting systems exhibit a similar behavior in the PAH~6.2/7.7 plane. The photoelectric efficiency is lower in the nuclear regions and increases toward larger galactocentric distances, even when the PAH~6.2/7.7 ratio remains relatively unchanged. In NGC~7674, the PAH~6.2/7.7 ratio increases with galactocentric distance, coinciding with elevated photoelectric-efficiency values. This behavior suggests that, while variations in PAH grain size may contribute to the photoelectric heating budget in the most AGN-dominated systems, the PAH grain size does not appear to be correlated with the photoelectric efficiency observed in galaxies hosting more moderate AGNs. 

Taken together, these results show that in both starburst- and AGN-hosting galaxies the photoelectric efficiency increases with galactocentric distance as the radiation field weakens, with PAH ionization state tracing this trend more directly than PAH grain size, particularly in starburst galaxies.
These trends may have implications for star formation in the early Universe, where galaxies are observed to sustain high star-formation rates relative to their gas reservoirs, implying elevated star-formation efficiencies \citep{McKinney20, McKinney21}. In intense radiation fields, a reduced photoelectric heating efficiency—associated with increasingly charged dust grains and PAHs—can weaken radiative feedback by lowering the effectiveness of gas heating, providing a pathway for maintaining high star-formation efficiencies in high-redshift systems, where radiation field densities are typically an order of magnitude or more higher than in local galaxies of similar stellar mass.
\vspace{-0.1cm}
\section{Summary and conclusions}
\label{conclusions}
We present a spatially resolved analysis of PAH properties and photoelectric heating in a representative subsample of 67 nearby LIRGs drawn from the GOALS sample. Using mid-IR spectral mapping from \textit{Spitzer}/IRS, complemented by \textit{Herschel}/PACS measurements of the main PDR cooling lines used to construct a photoelectric-efficiency proxy, we extracted spectra in concentric annuli and in matched square-aperture grids to trace how PAH band ratios and gas-heating efficiency vary from the nuclear regions into the extended disks on kiloparsec scales. Our analysis focuses on environments beyond the spatial coverage typically accessible to a single pointing of JWST IFU observations, providing constraints on the extended ISM where the imprint of central activity and feedback can persist. The main results of this work can be summarized as follows:
\begin{enumerate}
    \vspace{-0.1cm}
    \item We measured spatially resolved PAH band ratios (PAH~11.3/7.7 and PAH~6.2/7.7) across LIRGs and used them to trace changes in PAH ionization state and characteristic grain size with galactocentric distance. Both starburst- and AGN-hosting systems exhibit radial variations in these diagnostics, which demonstrates that PAH properties are not uniform across LIRGs and that integrated measurements can obscure environmentally driven trends. \\ 
    \vspace{-0.2cm}
    \item In starburst-dominated galaxies, the PAH ionization state shows a systematic radial behavior: the PAH~11.3/7.7 ratio increases from the nuclear regions out to $\sim$5~kpc, while PAH sizes remain constant. At intermediate radii (5--8~kpc), the PAH~11.3/7.7 ratio continues to rise, whereas the PAH size distribution begins to shift toward smaller grains. Beyond $\sim$8~kpc, the trend in PAH ionization reverses, with decreasing PAH~11.3/7.7 ratios indicating increasingly ionized PAHs, accompanied by a modest but statistically significant shift toward smaller PAH sizes at larger galactocentric distances. These trends are consistent with a radial decrease in radiation field intensity, which primarily regulates PAH sizes at large radii, supporting the scenario of the survival of larger PAHs in harsh radiation environments, while the PAH ionization state shows a weaker dependence on the local radiation field beyond the inner regions. \\ 
    \vspace{-0.2cm}
    \item Active galactic nucleus-hosting galaxies exhibit a broader range of PAH~6.2/7.7 and PAH~11.3/7.7 ratios than purely starburst galaxies. The nuclear regions of galaxies with strong AGNs are distinctly offset from the starburst locus, showing systematically higher PAH~11.3/7.7, lower PAH~6.2/7.7, and higher [NeIII]$_{15}$/[NeII]$_{13}$ ratios, which is consistent with the selective processing of small, ionized PAHs and the survival of larger, more neutral grains. These offsets persist into the extended regions at intermediate radii (up to $\sim$5~kpc), while over the same radial range the PAH size distribution is dominated by smaller grains. This suggests that AGN-related modifications to the radiation field and PAH emission are not confined to the nucleus but extend over several kiloparsecs. 
    In contrast, systems with more moderate AGN fractions overlap more closely with the starburst population, particularly at larger radii, which also shows a declining intensity of the radiation field. \\ 
    \vspace{-0.2cm}
    \item We used the commonly adopted proxy for the photoelectric efficiency,
    $([{\rm C\,II}]_{\rm PDR} + [{\rm O\,I}]) / L_{\rm PAH}$, and found that both starburst- and AGN-hosting galaxies display a systematic increase in photoelectric efficiency with galactocentric distance. This behavior indicates that gas heating is suppressed in the inner regions of LIRGs and becomes progressively more efficient in the extended disks. \\
    \vspace{-0.2cm}
    \item In starburst-dominated systems, the dominant factor regulating the photoelectric efficiency is the decrease of the radiation field intensity with galactocentric distance. The radially increasing $\epsilon_{\rm PE}$ correlates most directly with PAH~11.3/7.7, which is consistent with a scenario in which the weakening radiation field toward larger radii reduces positive grain charging and increases the photoelectric yield. In contrast, PAH~6.2/7.7 remains constant over the entire disk while the efficiency increases, which indicates that changes in PAH size are not the primary driver of the efficiency gradient over most radii. \\
    \vspace{-0.2cm}
    \item In AGN-hosting systems, the photoelectric efficiency also rises with radius, which supports the hypothesis that the large-scale efficiency increase is mainly governed by the decline of the radiation-field intensity with galactocentric distance. The relation between $\epsilon_{\rm PE}$ and PAH ionization/size may be more complex and seems to diverge from the trend observed in starburst galaxies. Nevertheless, any interpretation should be treated with caution, especially in the most AGN-dominated source, given the limited statistical constraints. 
\end{enumerate}
\vspace{-0.1cm}
Overall, our results demonstrate that PAH properties and photoelectric heating efficiency vary significantly within LIRGs and that resolved measurements are essential for linking the PAH population to the thermal balance of the neutral ISM. Extending this approach with higher spatial resolution, a larger sample of LIRGs that host powerful AGNs to improve statistical robustness, and additional multiwavelength constraints will enable a more complete view of how star formation and nuclear activity regulate dust processing and gas heating across the disks of LIRGs.
Our results pave the way for future high-resolution studies (e.g., with JWST/MIRI and ALMA) to further dissect the multiphase ISM in these extreme environments. 

\begin{acknowledgements}

We thank the referee for the thoughtful suggestions. CML and MSG acknowledge the research project was supported by the Hellenic Foundation for Research and Innovation (HFRI) under the "2nd Call for HFRI Research Projects to support Faculty Members $\&$ Researchers" (Project Number: 03382).
CML wishes to thank the "Summer School for Astrostatistics in Crete" for providing training on the statistical methods adopted in this work. 
MB acknowledges support from the Juan de La Cierva scholarship with reference JDC2023-052684-I, funded by MICIU/AEI/10.13039/501100011033 y por el FSE+ and from the Agencia Estatal de Investigacion of the Ministerio de Ciencia, Innovacion y Universidades (MCIU/AEI) under the grant ``Tracking active galactic nuclei feedback from parsec to kiloparsec scales'', with reference PID2022${-}$141105NB${-}$I00 and the European Regional Development Fund (ERDF).
DBS gratefully acknowledges support from NSF Grant 2407752.
The National Radio Astronomy Observatory and Green Bank Observatory are facilities of the U.S. National Science Foundation operated under cooperative agreement by Associated Universities, Inc.
IS acknowledges fundings from the European Research Council (ERC)
DistantDust (Grant No.101117541) and the Atracc\'{i}on de Talento Grant
No.2022-T1/TIC-20472 of the Comunidad de Madrid, Spain. L.A.B. acknowledges support from the Dutch Research Council (NWO) under grant VI.Veni.242.055 (\url{https://doi.org/10.61686/LAJVP77714}).

\end{acknowledgements}

\bibliographystyle{aa}
\bibliography{biblio}

\begin{appendix}
\clearpage   

\onecolumn       

\section{GOALS sample properties}

\begin{table}[h]                    
\centering
\footnotesize
\caption{GOALS sample properties.}
\label{tab:goals_sample}
\begin{tabular}{lcccccccc}
\hline\hline
Galaxy Name$^{(1)}$ & RA$^{(2)}$ & Dec$^{(3)}$ & Dist.$^{(4)}$ & Merger$^{(5)}$ & $L_{\rm IR}^{(6)}$ & AGN$^{(7)}$ & $M_*^{(8)}$ & SFR$^{(9)}$ \\
 & (deg) & (deg) & (Mpc) & Stage & ($L_\odot$) & Fraction & ($M_\odot$) & ($M_\odot$ yr$^{-1}$) \\
\hline
NGC0034 & 2.777 & $-12.108$ & $84.1$ & 4 & $(3.2 \pm 0.1) \times 10^{11}$ & $0.200 \pm 0.109$ & $(8.9 \pm 1.8) \times 10^{10}$ & $45.540 \pm 1.503$ \\
Arp256 & 4.712 & $-10.377$ & $117.5$ & 2 & $(4.7 \pm 0.1) \times 10^{11}$ & $0.159 \pm 0.078$ & $(5.2 \pm 1.0) \times 10^{10}$ & $65.270 \pm 3.433$ \\
arp236\_N & 16.948 & $-17.507$ & $85.5$ & 2 & $(5.4 \pm 0.1) \times 10^{11}$ & $0.322 \pm 0.133$ & $(1.7 \pm 0.3) \times 10^{11}$ & $70.920 \pm 3.117$ \\
ESO244-G012 & 19.535 & $-44.462$ & $91.5$ & 2 & $(2.5 \pm 0.1) \times 10^{11}$ & $0.159 \pm 0.066$ & $(1.4 \pm 0.3) \times 10^{11}$ & $35.150 \pm 1.559$ \\
IIIZw35 & 26.127 & 17.103 & $119.0$ & 2 & $(4.1 \pm 0.0) \times 10^{11}$ & $0.150 \pm 0.055$ & $(6.3 \pm 1.3) \times 10^{10}$ & $60.410 \pm 0.732$ \\
NGC0828 & 32.540 & 39.190 & $76.3$ & 4 & $(2.2 \pm 0.1) \times 10^{11}$ & $0.250 \pm 0.107$ & $(2.0 \pm 0.4) \times 10^{11}$ & $28.720 \pm 2.595$ \\
NGC1572 & 65.678 & $-40.601$ & $88.6$ & 0 & $(2.0 \pm 0.1) \times 10^{11}$ & $0.287 \pm 0.112$ & $(2.0 \pm 0.4) \times 10^{11}$ & $27.200 \pm 1.670$ \\
UGC03351 & 86.450 & 58.701 & $65.8$ & 1 & $(2.2 \pm 0.1) \times 10^{11}$ & $0.246 \pm 0.095$ & $(9.7 \pm 1.9) \times 10^{10}$ & $30.060 \pm 1.493$ \\
UGC03410\_E & 93.623 & 80.450 & $59.7$ & 1 & $(9.2 \pm 1.5) \times 10^{10}$ & $0.265 \pm 0.115$ & $(8.6 \pm 1.7) \times 10^{10}$ & $12.150 \pm 2.188$ \\
UGC03410\_W & 93.491 & 80.476 & $59.7$ & 1 & $(1.9 \pm 0.4) \times 10^{10}$ & $0.311 \pm 0.142$ & $(3.7 \pm 0.7) \times 10^{10}$ & $2.353 \pm 0.616$ \\
ESO557-G002\_N & 97.947 & $-17.621$ & $93.6$ & 1 & $(1.6 \pm 0.1) \times 10^{11}$ & $0.122 \pm 0.073$ & $(2.5 \pm 0.5) \times 10^{10}$ & $23.760 \pm 1.087$ \\
ESO557-G002\_S & 97.947 & $-17.621$ & $93.6$ & 1 & $(3.1 \pm 0.3) \times 10^{10}$ & $0.198 \pm 0.094$ & $(2.2 \pm 0.4) \times 10^{10}$ & $4.190 \pm 0.448$ \\
UGC03608 & 98.917 & 68.936 & $94.3$ & 2 & $(2.3 \pm 0.1) \times 10^{11}$ & $0.172 \pm 0.060$ & $(9.7 \pm 1.9) \times 10^{10}$ & $32.570 \pm 1.552$ \\
NGC2342 & 107.237 & 20.637 & $78.0$ & 1 & $(1.9 \pm 0.3) \times 10^{11}$ & $0.204 \pm 0.080$ & $(1.3 \pm 0.3) \times 10^{11}$ & $26.120 \pm 4.599$ \\
NGC2388 & 112.525 & 33.835 & $62.1$ & 1 & $(2.1 \pm 0.2) \times 10^{11}$ & $0.104 \pm 0.079$ & $(7.7 \pm 1.5) \times 10^{10}$ & $30.870 \pm 3.435$ \\
MCG+02-20-003\_S & 116.350 & 10.308 & $72.8$ & 1 & $(1.7 \pm 0.0) \times 10^{10}$ & $0.315 \pm 0.108$ & $(2.4 \pm 0.5) \times 10^{10}$ & $2.154 \pm 0.047$ \\
NGC2623 & 129.600 & 25.757 & $84.1$ & 4 & $(4.5 \pm 0.0) \times 10^{11}$ & $0.259 \pm 0.127$ & $(6.4 \pm 1.3) \times 10^{10}$ & $60.610 \pm 1.757$ \\
Arp303\_N & 139.225 & 36.943 & $92.9$ & 1 & $(6.9 \pm 1.0) \times 10^{10}$ & $0.306 \pm 0.125$ & $(1.2 \pm 0.2) \times 10^{11}$ & $8.747 \pm 1.516$ \\
Arp303\_S & 139.242 & 36.911 & $92.9$ & 1 & $(8.7 \pm 0.8) \times 10^{10}$ & $0.236 \pm 0.096$ & $(5.7 \pm 1.1) \times 10^{10}$ & $11.720 \pm 1.292$ \\
NGC3110 & 151.246 & $-6.476$ & $79.5$ & 1 & $(2.2 \pm 0.1) \times 10^{11}$ & $0.229 \pm 0.096$ & $(1.3 \pm 0.3) \times 10^{11}$ & $29.490 \pm 2.061$ \\
NGC3256 & 156.963 & $-43.904$ & $38.9$ & 4 & $(4.4 \pm 0.0) \times 10^{11}$ & $0.159 \pm 0.060$ & $(1.1 \pm 0.2) \times 10^{11}$ & $61.690 \pm 1.329$ \\
ESO264-G036 & 161.596 & $-43.471$ & $100.0$ & 0 & $(1.8 \pm 0.2) \times 10^{11}$ & $0.321 \pm 0.114$ & $(2.5 \pm 0.5) \times 10^{11}$ & $23.140 \pm 3.435$ \\
ESO320-G030 & 163.633 & $-39.460$ & $41.2$ & 0 & $(1.5 \pm 0.0) \times 10^{11}$ & $0.165 \pm 0.103$ & $(4.6 \pm 0.9) \times 10^{10}$ & $22.000 \pm 0.439$ \\
ESO440-IG058\_S & 169.204 & $-31.653$ & $112.0$ & 2 & $(2.6 \pm 0.0) \times 10^{11}$ & $0.179 \pm 0.080$ & $(5.0 \pm 1.0) \times 10^{10}$ & $37.000 \pm 1.379$ \\
ESO440-IG058\_N & 169.221 & $-31.627$ & $112.0$ & 2 & $(3.1 \pm 0.1) \times 10^{10}$ & $0.176 \pm 0.095$ & $(8.0 \pm 1.6) \times 10^{10}$ & $4.387 \pm 0.144$ \\
MCG-02-33-098\_E & 193.221 & $-13.194$ & $78.7$ & 2 & $(8.7 \pm 0.1) \times 10^{10}$ & $0.245 \pm 0.092$ & $(1.9 \pm 0.4) \times 10^{10}$ & $11.590 \pm 0.797$ \\
MCG-02-33-098\_W & 193.242 & $-13.185$ & $78.7$ & 2 & $(3.0 \pm 0.1) \times 10^{10}$ & $0.155 \pm 0.080$ & $(1.8 \pm 0.4) \times 10^{10}$ & $4.237 \pm 0.173$ \\
NGC5135 & 201.396 & $-29.835$ & $60.9$ & 0 & $(2.2 \pm 0.0) \times 10^{11}$ & $0.356 \pm 0.084$ & $(1.3 \pm 0.3) \times 10^{11}$ & $24.880 \pm 2.129$ \\
ESO173-G015 & 203.017 & $-57.428$ & $34.0$ & 0 & $(2.5 \pm 0.0) \times 10^{11}$ & $0.156 \pm 0.089$ & $(3.3 \pm 0.7) \times 10^{10}$ & $36.720 \pm 0.642$ \\
NGC5256\_S & 204.567 & 48.276 & $129.0$ & 2 & $(3.2 \pm 0.1) \times 10^{11}$ & $0.332 \pm 0.059$ & $(6.3 \pm 1.3) \times 10^{10}$ & $36.160 \pm 3.434$ \\
Arp240\_E & 208.396 & 33.004 & $108.5$ & 2 & $(2.5 \pm 0.7) \times 10^{11}$ & $0.254 \pm 0.087$ & $(1.6 \pm 0.3) \times 10^{11}$ & $33.330 \pm 8.866$ \\
Arp240\_W & 208.379 & 33.015 & $108.5$ & 2 & $(2.7 \pm 0.7) \times 10^{11}$ & $0.190 \pm 0.077$ & $(1.4 \pm 0.3) \times 10^{11}$ & $36.740 \pm 9.361$ \\
ESO221-IG010 & 208.946 & $-49.367$ & $62.9$ & 0 & $(1.7 \pm 0.0) \times 10^{11}$ & $0.177 \pm 0.081$ & $(1.3 \pm 0.3) \times 10^{11}$ & $23.160 \pm 1.054$ \\
NGC5395 & 209.121 & 37.433 & $58.7$ & 3 & $(7.7 \pm 0.2) \times 10^{10}$ & $0.805 \pm 0.043$ & $(1.7 \pm 0.3) \times 10^{11}$ & $5.889 \pm 0.208$ \\
NGC5653 & 217.329 & 31.215 & $60.2$ & 0 & $(4.6 \pm 1.2) \times 10^{10}$ & $0.236 \pm 0.095$ & $(9.3 \pm 1.9) \times 10^{10}$ & $6.253 \pm 1.671$ \\
NGC5734\_N & 220.883 & $-20.865$ & $67.1$ & 1 & $(1.1 \pm 0.1) \times 10^{11}$ & $0.330 \pm 0.123$ & $(1.2 \pm 0.2) \times 10^{11}$ & $13.720 \pm 1.333$ \\
NGC5734\_S & 220.892 & $-20.885$ & $67.1$ & 1 & $(6.8 \pm 0.3) \times 10^{10}$ & $0.335 \pm 0.119$ & $(6.6 \pm 1.3) \times 10^{10}$ & $8.426 \pm 0.791$ \\
Zw049.057 & 225.921 & 16.443 & $65.4$ & 0 & $(2.5 \pm 0.0) \times 10^{11}$ & $0.182 \pm 0.107$ & $(2.0 \pm 0.4) \times 10^{10}$ & $35.470 \pm 0.961$ \\
NGC5990 & 236.579 & $-25.643$ & $64.4$ & 1 & $(1.6 \pm 0.0) \times 10^{11}$ & $0.485 \pm 0.161$ & $(1.4 \pm 0.3) \times 10^{11}$ & $16.630 \pm 1.715$ \\
NGC6156 & 245.946 & $-60.684$ & $48.0$ & 0 & $(1.1 \pm 0.0) \times 10^{11}$ & $0.318 \pm 0.100$ & $(1.0 \pm 0.2) \times 10^{11}$ & $14.020 \pm 0.849$ \\
NGC6240 & 253.246 & 2.401 & $116.0$ & 3 & $(9.4 \pm 0.4) \times 10^{11}$ & $0.289 \pm 0.109$ & $(3.9 \pm 0.8) \times 10^{11}$ & $123.200 \pm 5.817$ \\
NGC6286\_S & 255.346 & 58.947 & $85.7$ & 2 & $(2.1 \pm 0.5) \times 10^{11}$ & $0.253 \pm 0.099$ & $(1.3 \pm 0.3) \times 10^{11}$ & $28.020 \pm 7.329$ \\
NGC6286\_N & 255.358 & 58.963 & $85.7$ & 2 & $(7.2 \pm 1.8) \times 10^{10}$ & $0.201 \pm 0.088$ & $(3.8 \pm 0.8) \times 10^{10}$ & $9.884 \pm 2.554$ \\
IRASF17138-1017 & 259.125 & $-10.317$ & $84.0$ & 4 & $(3.4 \pm 0.1) \times 10^{11}$ & $0.182 \pm 0.082$ & $(1.5 \pm 0.3) \times 10^{11}$ & $46.620 \pm 2.055$ \\
ESO138-G027 & 262.404 & $-59.215$ & $98.3$ & 0 & $(2.8 \pm 0.1) \times 10^{11}$ & $0.206 \pm 0.078$ & $(7.1 \pm 1.4) \times 10^{10}$ & $38.210 \pm 1.110$ \\
NGC6621 & 275.746 & 68.360 & $94.3$ & 2 & $(2.1 \pm 0.1) \times 10^{11}$ & $0.256 \pm 0.100$ & $(1.3 \pm 0.3) \times 10^{11}$ & $27.570 \pm 2.480$ \\
IC4687 & 275.804 & $-57.724$ & $81.9$ & 2 & $(2.3 \pm 0.0) \times 10^{11}$ & $0.188 \pm 0.088$ & $(8.7 \pm 1.7) \times 10^{10}$ & $31.170 \pm 2.103$ \\
IC4734 & 281.821 & $-49.752$ & $73.4$ & 0 & $(2.7 \pm 0.0) \times 10^{11}$ & $0.216 \pm 0.089$ & $(9.1 \pm 1.8) \times 10^{10}$ & $36.940 \pm 0.919$ \\

\hline
\end{tabular}
\end{table}

\begin{table}[h]                    
\centering
\footnotesize
\caption*{continued.}
\begin{tabular}{lcccccccc}
\hline\hline
Galaxy Name$^{(1)}$ & RA$^{(2)}$ & Dec$^{(3)}$ & Dist.$^{(4)}$ & Merger$^{(5)}$ & $L_{\rm IR}^{(6)}$ & AGN$^{(7)}$ & $M_*^{(8)}$ & SFR$^{(9)}$ \\
 & (deg) & (deg) & (Mpc) & Stage & ($L_\odot$) & Fraction & ($M_\odot$) & ($M_\odot$ yr$^{-1}$) \\
\hline
NGC6701 & 282.525 & $-60.606$ & $62.4$ & 0 & $(1.1 \pm 0.0) \times 10^{11}$ & $0.229 \pm 0.087$ & $(9.8 \pm 2.0) \times 10^{10}$ & $15.480 \pm 0.733$ \\
VV414 & 286.258 & 55.388 & $113.0$ & 2 & $(1.6 \pm 0.1) \times 10^{11}$ & $0.207 \pm 0.094$ & $(1.3 \pm 0.3) \times 10^{11}$ & $21.940 \pm 1.314$ \\
VV414\_E & 286.342 & 55.388 & $113.0$ & 2 & $(9.3 \pm 1.2) \times 10^{10}$ & $0.364 \pm 0.126$ & $(1.1 \pm 0.2) \times 10^{11}$ & $11.610 \pm 1.677$ \\
NGC6926 & 306.321 & 57.050 & $89.1$ & 1 & $(7.8 \pm 0.8) \times 10^{10}$ & $0.422 \pm 0.103$ & $(1.9 \pm 0.4) \times 10^{11}$ & $8.247 \pm 1.070$ \\
CGCG448-020\_W & 318.067 & 38.515 & $161.0$ & 3 & $(7.3 \pm 0.1) \times 10^{11}$ & $0.142 \pm 0.052$ & $(2.7 \pm 0.5) \times 10^{10}$ & $103.300 \pm 1.880$ \\
CGCG448-020\_E & 318.092 & 38.520 & $161.0$ & 2 & $(5.9 \pm 1.9) \times 10^{11}$ & $0.200 \pm 0.109$ & $(3.8 \pm 0.8) \times 10^{10}$ & $82.360 \pm 2.682$ \\
ESO343-IG013\_S & 318.583 & $-38.413$ & $85.8$ & 3 & $(4.4 \pm 0.1) \times 10^{10}$ & $0.171 \pm 0.092$ & $(6.1 \pm 1.2) \times 10^{10}$ & $6.253 \pm 0.171$ \\
ESO343-IG013\_N & 318.592 & $-38.396$ & $85.8$ & 3 & $(8.3 \pm 0.1) \times 10^{10}$ & $0.244 \pm 0.087$ & $(1.5 \pm 0.3) \times 10^{10}$ & $11.230 \pm 0.537$ \\
NGC7130 & 327.825 & $-34.951$ & $72.7$ & 0 & $(2.5 \pm 0.0) \times 10^{11}$ & $0.391 \pm 0.115$ & $(1.4 \pm 0.3) \times 10^{11}$ & $29.470 \pm 1.530$ \\
IC5179 & 333.071 & $-36.839$ & $51.4$ & 0 & $(1.5 \pm 0.0) \times 10^{11}$ & $0.223 \pm 0.096$ & $(1.2 \pm 0.2) \times 10^{11}$ & $19.960 \pm 1.578$ \\
NGC7469 & 345.817 & 8.873 & $70.8$ & 2 & $(3.8 \pm 0.1) \times 10^{11}$ & $0.435 \pm 0.142$ & $(1.9 \pm 0.4) \times 10^{11}$ & $42.670 \pm 3.387$ \\
NGC7552 & 349.021 & $-42.588$ & $23.5$ & 0 & $(1.4 \pm 0.0) \times 10^{11}$ & $0.216 \pm 0.082$ & $(7.6 \pm 1.5) \times 10^{10}$ & $18.580 \pm 0.948$ \\
NGC7591 & 349.613 & 6.590 & $71.4$ & 0 & $(1.4 \pm 0.0) \times 10^{11}$ & $0.298 \pm 0.101$ & $(9.8 \pm 2.0) \times 10^{10}$ & $18.430 \pm 0.711$ \\
ngc7592\_E & 349.846 & $-4.424$ & $106.0$ & 2 & $(1.2 \pm 0.0) \times 10^{11}$ & $0.152 \pm 0.077$ & $(7.2 \pm 1.4) \times 10^{10}$ & $17.500 \pm 0.423$ \\
NGC7592\_W & 349.821 & $-4.424$ & $106.0$ & 2 & $(7.5 \pm 0.2) \times 10^{10}$ & $0.395 \pm 0.132$ & $(3.9 \pm 0.8) \times 10^{10}$ & $8.850 \pm 0.688$ \\
NGC7674 & 351.983 & 8.779 & $125.0$ & 2 & $(2.6 \pm 0.2) \times 10^{11}$ & $0.798 \pm 0.067$ & $(2.5 \pm 0.5) \times 10^{11}$ & $10.380 \pm 2.983$ \\
Arp86\_N & 352.600 & 26.393 & $73.6$ & 2 & $(1.2 \pm 0.4) \times 10^{10}$ & $0.359 \pm 0.138$ & $(2.6 \pm 0.5) \times 10^{11}$ & $1.524 \pm 0.556$ \\
Arp86\_S & 352.588 & 26.378 & $73.6$ & 2 & $(5.6 \pm 2.0) \times 10^{10}$ & $0.184 \pm 0.091$ & $(2.4 \pm 0.5) \times 10^{10}$ & $7.655 \pm 2.784$ \\
NGC7771\_N & 358.646 & 20.112 & $61.2$ & 1 & $(2.5 \pm 0.2) \times 10^{11}$ & $0.224 \pm 0.100$ & $(2.4 \pm 0.5) \times 10^{11}$ & $34.130 \pm 3.037$ \\
\hline
\end{tabular}
\vspace{1em}
\tablefoot{Columns: (1) galaxy name; (2) right ascension and (3) declination in degrees from \citealt{Armus09}; (4) distance to the galaxies taken from \citealt{Armus09}; (5) merger stage as defined from \citealt{Sabrina13}; (6) total infrared luminosities from \citealt{Armus09}; (7) integrated MIR AGN fraction as defined in \citealt{Tanio17}, (8) total stellar mass from \citealt{Howell10} and (9) total SFR from \citealt{Howell10}.}
\end{table}


\newpage
\section{Stacked spectra}
\begin{figure}[!htbp] 
    \centering
    \includegraphics[width=\columnwidth]{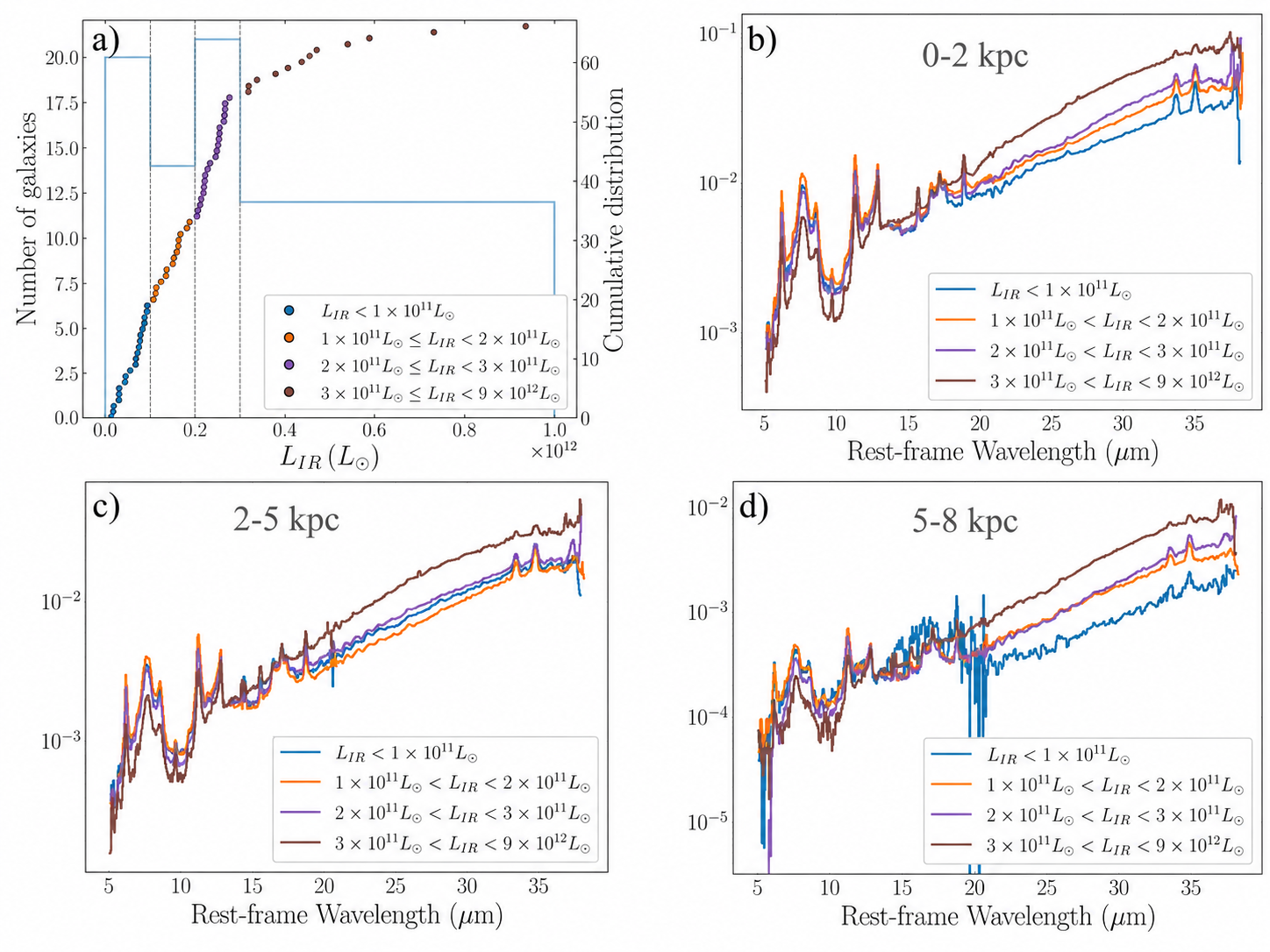}
    \caption{Same as Fig. \ref{fig:AGN_stacked} but the spectra are stacked according to their L$_{IR}$.}
    \label{fig:LIR_stacked}
\end{figure}

Here we show the galaxy sample spectra divided into four bins and stacked according to the integrated properties: L$_{IR}$, sSFR, and merger stage. 
Figure \ref{fig:LIR_stacked} displays the stacked spectra as a function of the $L_{IR}$. From the nuclear region up to the outermost one, the galaxies belonging to the last $L_{IR}$ bin (i.e., the most luminous galaxies in the sample) show weaker PAH lines and deeper silicate absorption, similar to what has been observed in local LIRGs \citep{Sabrina13}. This suggests that high L$_{\rm IR}$ galaxies have fundamentally different interstellar medium conditions, possibly including higher dust temperatures and/or altered dust grain properties throughout the galaxy. 

\begin{figure}[!htbp] 
    \centering
    \includegraphics[width=\columnwidth]{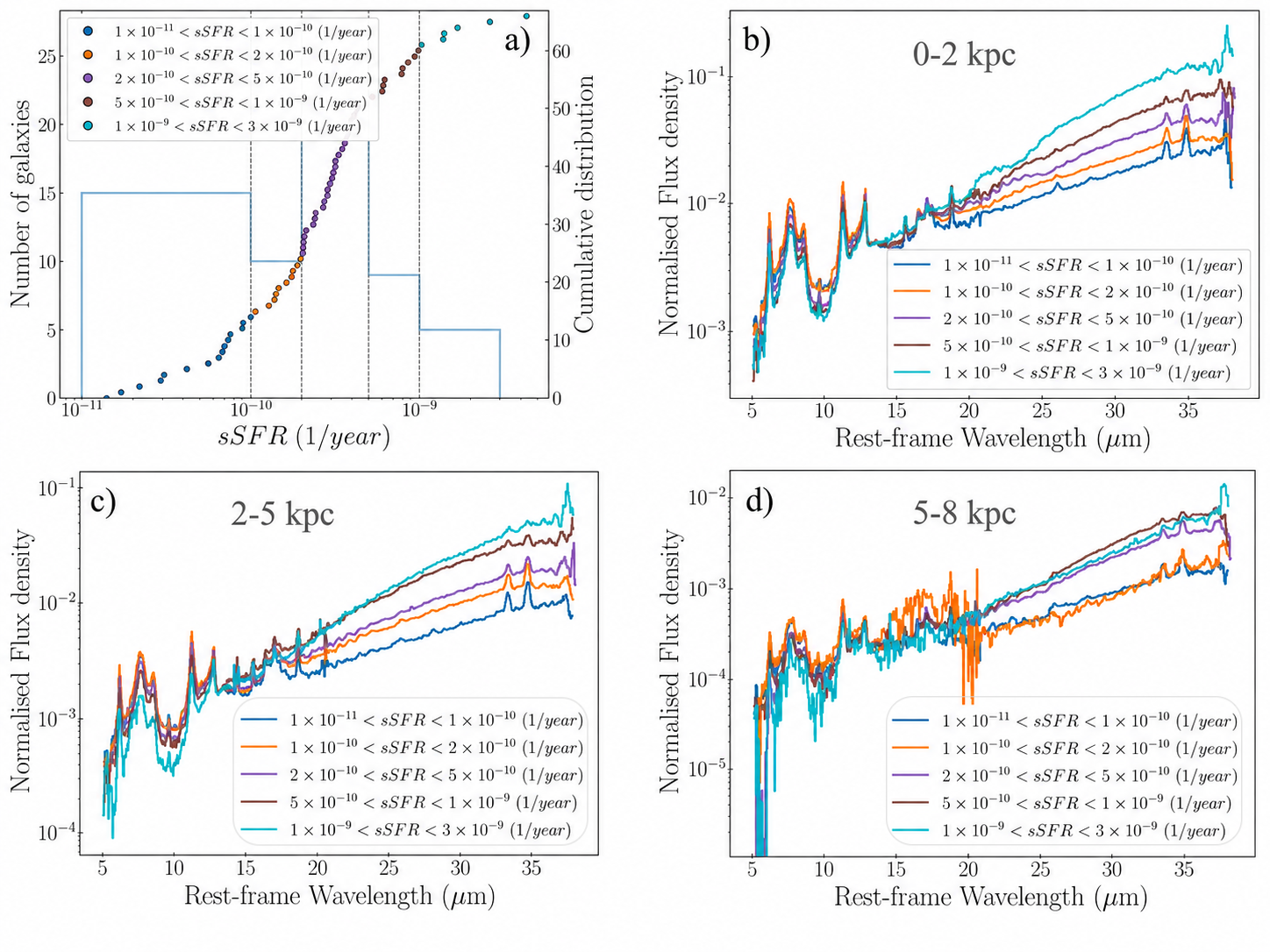}
    \caption{Same as Fig. \ref{fig:LIR_stacked} but the spectra are stacked according to the sSFR values.}
    \label{fig:ssfr_stacked}
\end{figure}

Figure \ref{fig:ssfr_stacked} presents the stacked spectra distributed into five bins based on the sSFR. As depicted, galaxies belonging to the last bins (i.e., the galaxies with the highest sSFR), have deeper silicate absorption. This could suggest that galaxies in active star-forming phases have fundamentally different dust characteristics, possibly due to recent gas accretion or merger activity, that manifest as enhanced absorption throughout the galaxy. 

\begin{figure}[!htbp] 
    \centering
    \includegraphics[width=\columnwidth]{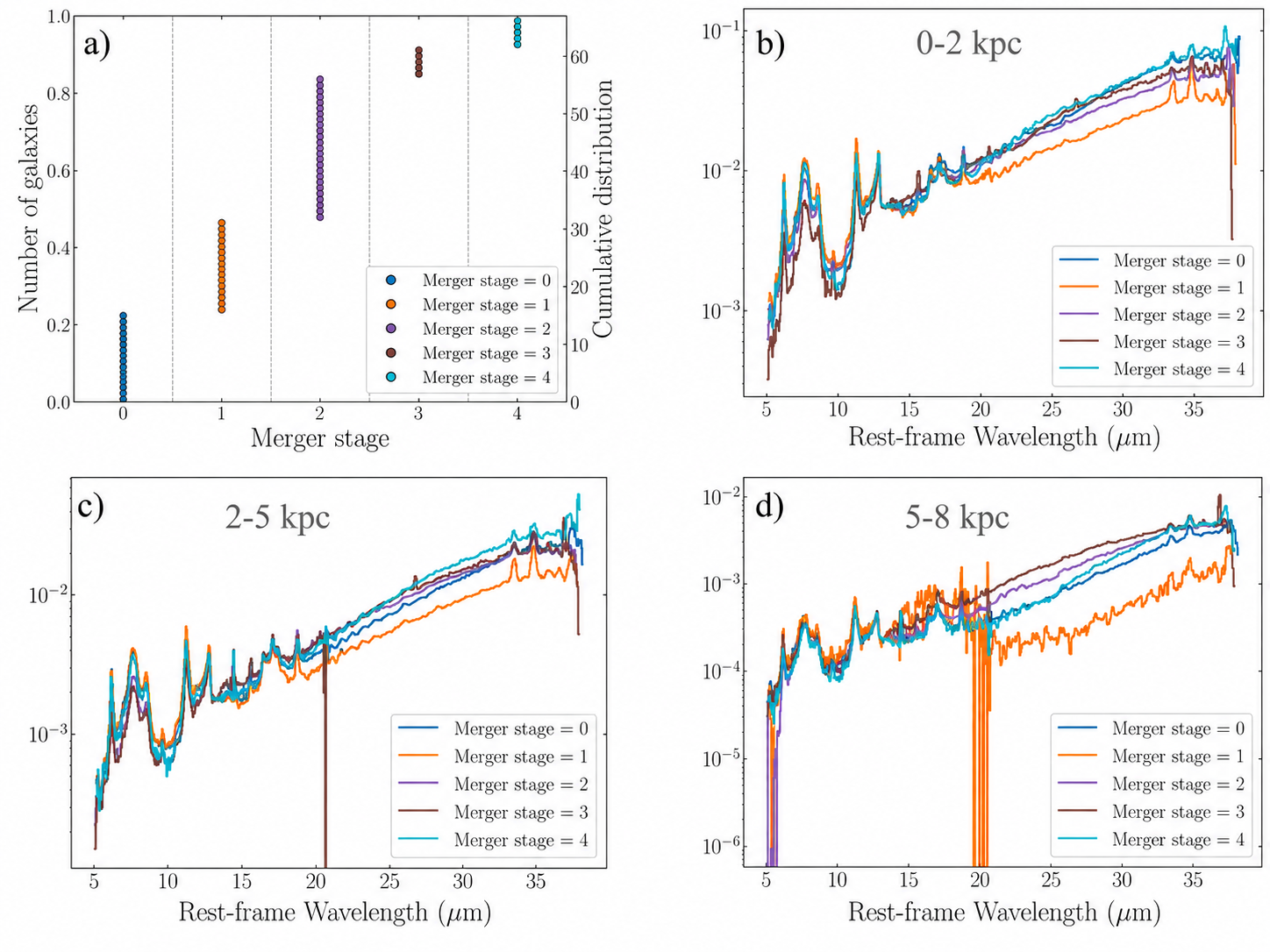}
    \caption{Same as Fig. \ref{fig:LIR_stacked} but the spectra are stacked according to the merger stage values.}
    \label{fig:merger_stacked}
\end{figure}

The spectra were also stacked based on the merger stage; however, the galaxies display similar spectral features regardless of their assigned bin 
(Fig.\ref{fig:merger_stacked}). 
\label{Appendix1}

\section{CAFE fits}
\label{cafe_fits}

\begin{figure}[] 
    \centering
    \includegraphics[width=0.7\linewidth]{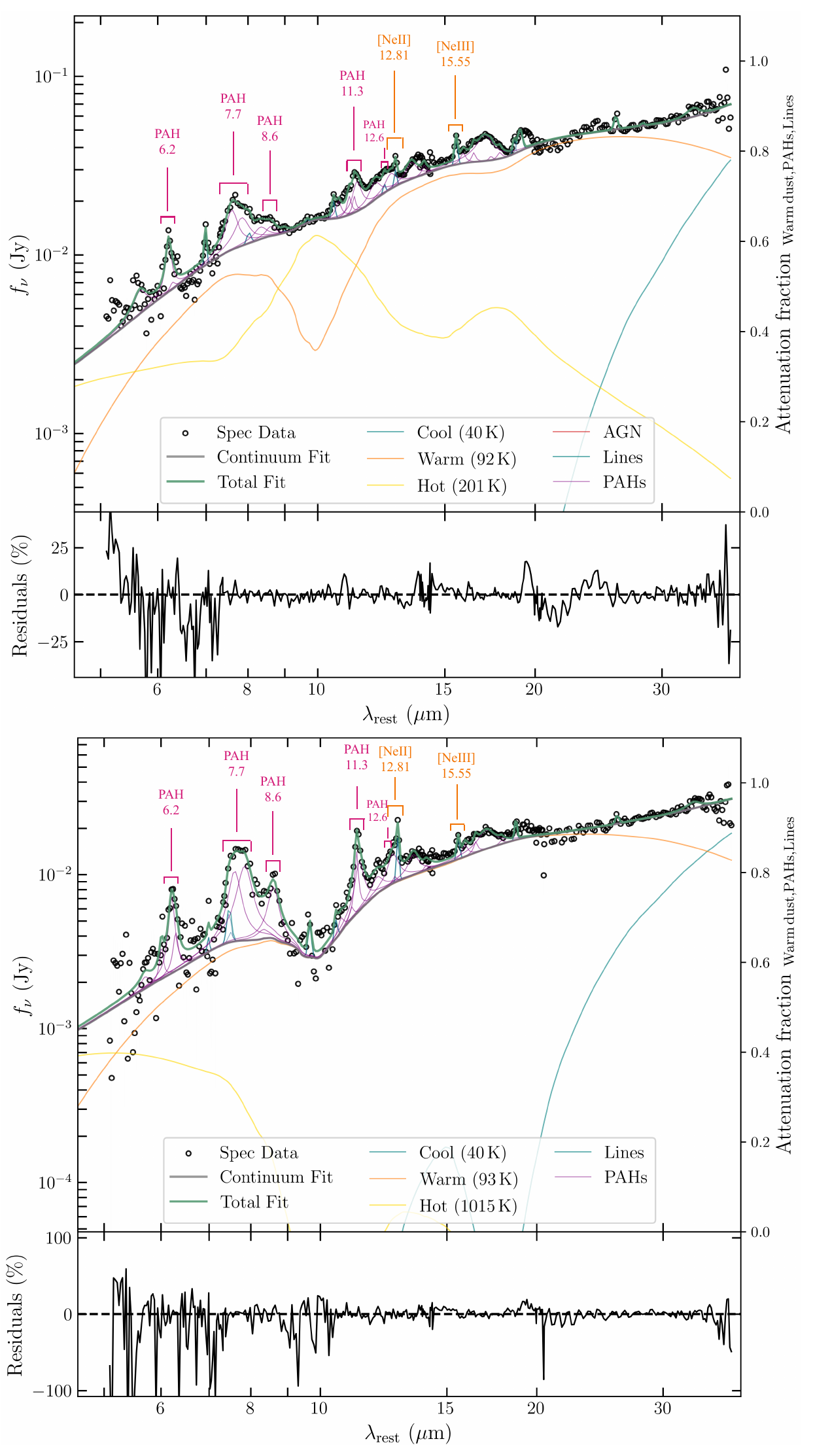}

    \captionof{figure}
    {Top: Mid-IR spectrum of the nuclear region of NGC~7674. 
    Bottom: Mid-IR spectrum of one of the most distant regions from the nucleus, shown for comparison. 
    The observed data are shown as black points, while the green line represents the total CAFE fit. 
    The gray curve shows the underlying continuum, composed of cool (cyan), warm (orange), and hot (yellow) dust components. 
    The purple curves indicate the polycyclic aromatic hydrocarbon (PAH) features, and the blue curves mark the atomic and molecular emission lines.}
    \label{fig:cafe_fits}
\end{figure}

As an illustrative example, Fig.~\ref{fig:cafe_fits} presents mid-IR spectra of NGC~7674 extracted from two distinct regions within the galaxy. The top panel shows the spectrum of the nuclear region, while the bottom panel displays the spectrum of one of the most distant regions from the nucleus included in our analysis. This comparison highlights the different spectral characteristics associated with nuclear and extended regions and illustrates how the CAFE fitting procedure provides an adequate description of the mid-IR emission in both cases. 

\FloatBarrier
\section{Radial profile of [Ne\,\textsc{iii}]\,15.56\,$\mu$m\,/\,[Ne\,\textsc{ii}]\,12.81\,$\mu$m ratio }
\begin{figure}[] 
\includegraphics[width=\columnwidth]{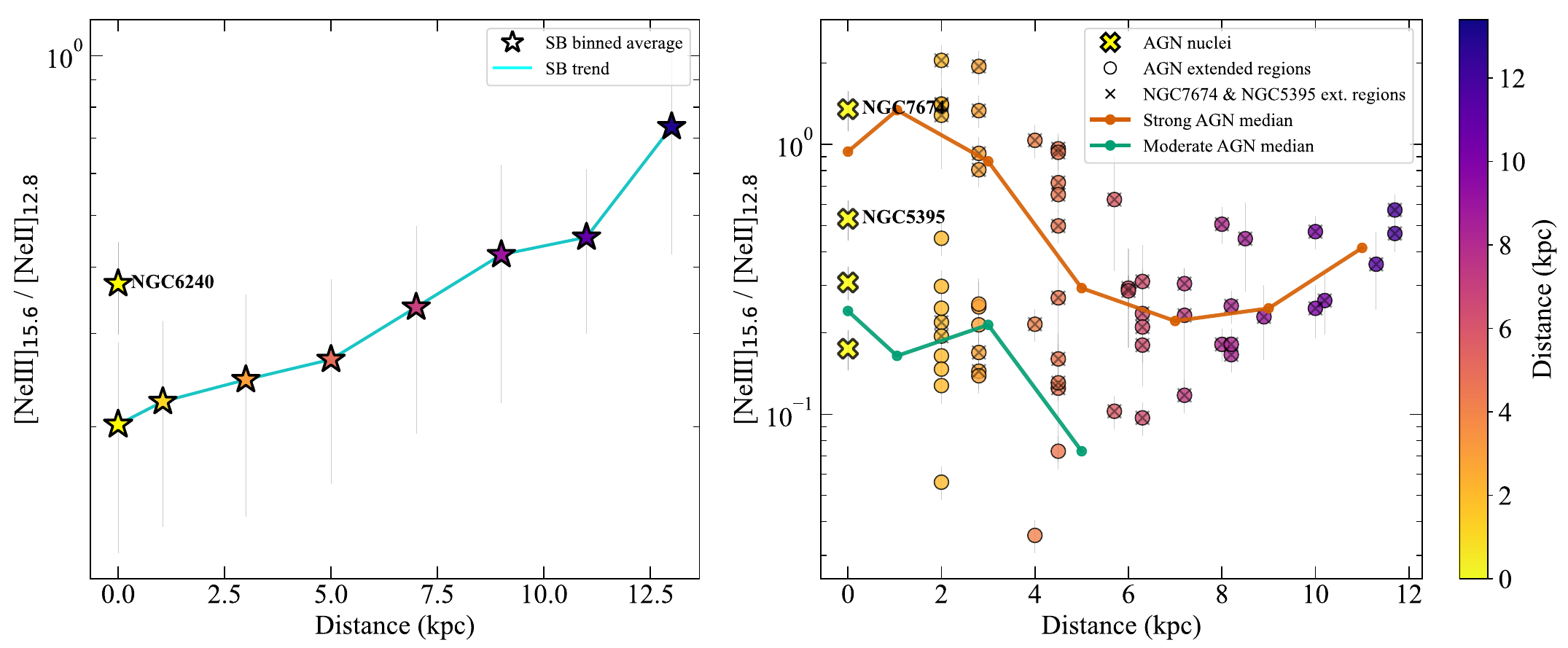}
\caption{Radial profiles of the [Ne\,\textsc{iii}]\,15.56\,$\mu$m\,/\,[Ne\,\textsc{ii}]\,12.81\,$\mu$m ratio as a function of projected galactocentric distance for the GOALS sample. Left: Starburst-dominated galaxies ($f_{\mathrm{AGN}}{<}0.4$) and  Right: AGN-dominated galaxies ($f_{\mathrm{AGN}}{\geq}0.4$). Symbols follow the same convention as in Fig.~\ref{fig:pah_ratio}.}
\label{fig:NeIII_NeII_vs_dist}
\end{figure}

\FloatBarrier

\section{Tables}
\label{PAH from Draine21}
\begin{figure}[!htbp] 
    \centering
    \includegraphics[width=\columnwidth]{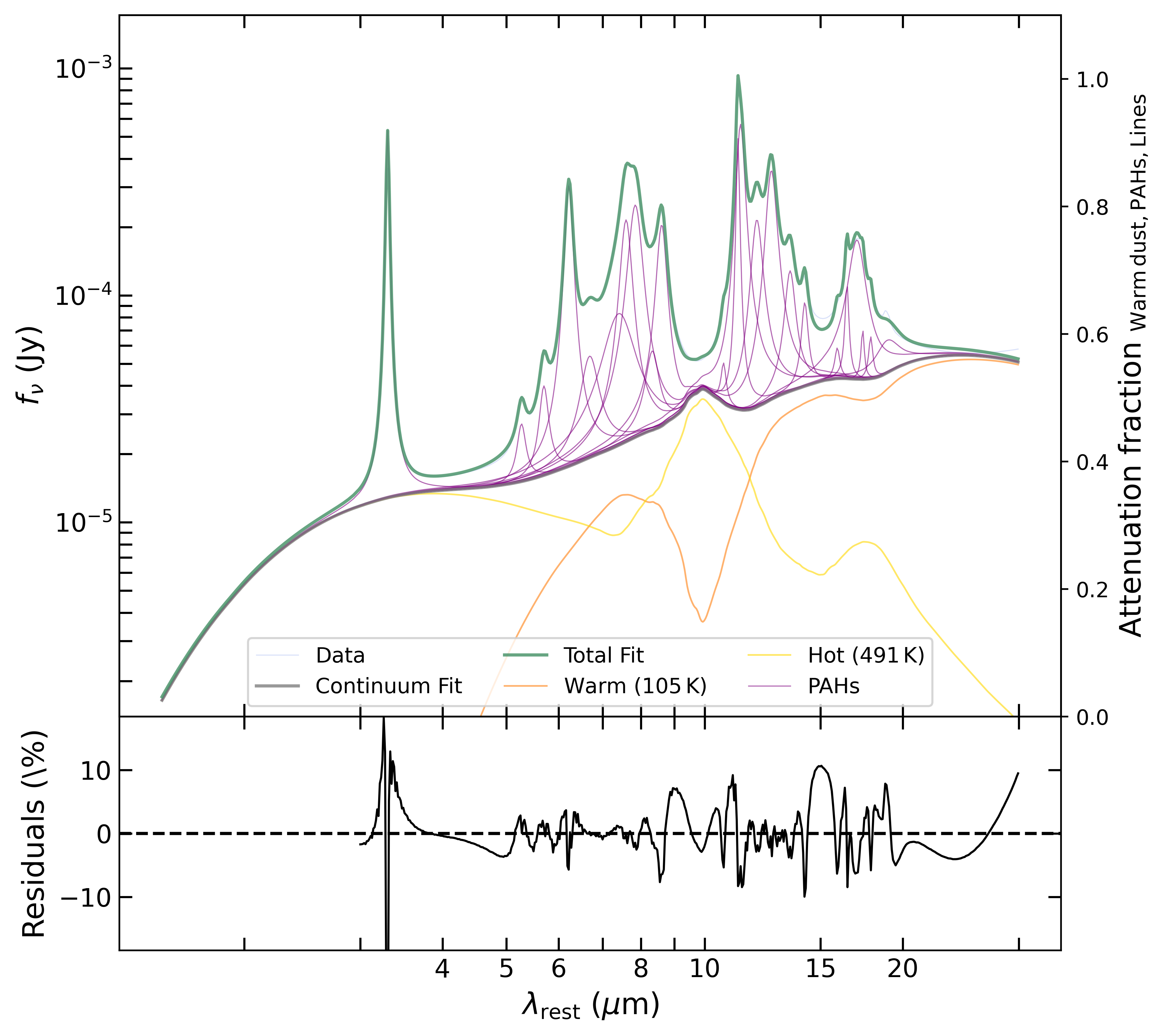}
    \caption{Example CAFE fit for neutral PAH with low ionization, small size and logU=0.}
    \label{fig:Draine_fit}
\end{figure}

In this section we provide tables of predicted PAH flux ratios for both neutral and ionized PAH populations, computed from the theoretical PAH emission spectra available in the library of \citet{Draine21}\footnote{\url{https://dataverse.harvard.edu/dataset.xhtml?persistentId=doi:10.7910/DVN/LPUHIQ}}. The spectra used as input are those generated assuming a BC03 single-age stellar population of 3~Myr with solar metallicity (Z=0.02) from \citet{BC03}, spanning the dimensionless intensity scaling parameter from logU=0 to logU=6, three ionization states (low, standard, high), and three grain size distributions (small, standard, large) for both neutral and ionized PAHs. We fitted each spectrum with CAFE to extract the individual PAH feature fluxes and subsequently computed the band ratios reported in Tables~\ref{tab:pah_ratios_neutral_U0.00}--\ref{tab:pah_ratios_ionized_U6.00} (an example of \texttt{CAFE} fit is reported in Fig.~\ref{fig:Draine_fit}).
We note that the diagnostic analysis presented in the main body of this work does not rely on these specific model grids. The parameter space occupied by the galaxies in our sample extends beyond the region covered by the grids parameterized by ionization state and grain size distribution. For this reason, we adopted the \citet{Draine21} models parameterized by the number of carbon atoms (N$_{\rm C}$), which span a broader range in PAH band ratio space and thus provide better coverage of the observed properties. The tables presented here are nonetheless included as a reference for the community, as they allow a direct mapping between the physical conditions of the radiation field and the resulting PAH emission ratios without the need to reprocess the theoretical spectra. It is worth noting that the spectra from \citep{Draine21} are fitted using the \texttt{CAFE} software. As a result, the PAH ratios reported in these tables may differ from those obtained using the “clipping” methodology adopted in \citep{Draine21}. We adopt this approach to ensure consistency, as the same fitting procedure is applied to both the observational data and the models.

\FloatBarrier

\begin{table*}[h!] 
\centering
\caption{Predicted PAH band ratios as a function of the number of carbon atoms ($N_{\rm C}$) from the \citet{Draine21} models, for neutral and ionized PAHs for logU=2}
\label{tab:pah_ratios_NC}
\begin{tabular}{ccccc}
\hline\hline
$N_{\rm C}$ & \multicolumn{2}{c}{Neutral} & \multicolumn{2}{c}{Ionized} \\
            & 6.2/7.7 & 11.3/7.7 & 6.2/7.7 & 11.3/7.7 \\
\hline
26   & 0.478 & 0.588 & 0.416 & 0.061 \\
37   & 0.457 & 0.518 & 0.401 & 0.054 \\
53   & 0.431 & 0.470 & 0.380 & 0.050 \\
74   & 0.405 & 0.454 & 0.351 & 0.049 \\
105  & 0.369 & 0.424 & 0.315 & 0.048 \\
148  & 0.338 & 0.483 & 0.291 & 0.059 \\
209  & 0.288 & 0.555 & 0.266 & 0.066 \\
296  & 0.248 & 0.663 & 0.240 & 0.092 \\
418  & 0.210 & 0.809 & 0.184 & 0.134 \\
590  & 0.175 & 1.022 & 0.155 & 0.179 \\
834  & 0.143 & 1.338 & 0.126 & 0.246 \\
1179 & 0.115 & 1.798 & 0.096 & 0.338 \\
\hline
\end{tabular}
\end{table*}

\begin{table*}[h!] 
\centering
\caption{Neutral PAH flux ratios for $\log U = 0.00$.}
\label{tab:pah_ratios_neutral_U0.00}
\begin{tabular}{llccccc}
\hline\hline
Size     & Ionization & 3.3/6.2 & 6.2/7.7 & 7.7/8.6 & 8.6/11.3 & 11.3/17 \\
\hline
small    & low       & 2.091 & 0.382 & 3.826 & 0.305 & 4.633 \\
standard & low       & 1.278 & 0.343 & 4.359 & 0.298 & 3.478 \\
large    & low       & 0.888 & 0.311 & 4.405 & 0.277 & 2.738 \\
small    & standard  & 2.061 & 0.383 & 3.929 & 0.312 & 4.760 \\
standard & standard  & 1.310 & 0.346 & 4.383 & 0.303 & 3.585 \\
large    & standard  & 0.919 & 0.314 & 4.428 & 0.281 & 2.816 \\
small    & high      & 2.057 & 0.384 & 4.002 & 0.318 & 4.883 \\
standard & high      & 1.349 & 0.349 & 4.399 & 0.307 & 3.689 \\
large    & high      & 0.956 & 0.318 & 4.446 & 0.286 & 2.899 \\
\hline
\end{tabular}
\end{table*}

\begin{table*}[h!] 
\centering
\caption{Neutral PAH flux ratios for $\log U = 1.00$.}
\label{tab:pah_ratios_neutral_U1.00}
\begin{tabular}{llccccc}
\hline\hline
Size     & Ionization & 3.3/6.2 & 6.2/7.7 & 7.7/8.6 & 8.6/11.3 & 11.3/17 \\
\hline
small    & low       & 2.018 & 0.382 & 3.939 & 0.306 & 4.401 \\
standard & low       & 1.284 & 0.344 & 4.365 & 0.295 & 3.367 \\
large    & low       & 0.894 & 0.312 & 4.403 & 0.273 & 2.667 \\
small    & standard  & 2.020 & 0.383 & 3.998 & 0.312 & 4.583 \\
standard & standard  & 1.315 & 0.347 & 4.387 & 0.300 & 3.496 \\
large    & standard  & 0.925 & 0.315 & 4.424 & 0.279 & 2.758 \\
small    & high      & 2.034 & 0.385 & 4.044 & 0.318 & 4.746 \\
standard & high      & 1.355 & 0.349 & 4.400 & 0.305 & 3.616 \\
large    & high      & 0.961 & 0.318 & 4.443 & 0.283 & 2.848 \\
\hline
\end{tabular}
\end{table*}

\begin{table*}[h!] 
\centering
\caption{Neutral PAH flux ratios for $\log U = 2.00$.}
\label{tab:pah_ratios_neutral_U2.00}
\begin{tabular}{llccccc}
\hline\hline
Size     & Ionization & 3.3/6.2 & 6.2/7.7 & 7.7/8.6 & 8.6/11.3 & 11.3/17 \\
\hline
small    & low       & 1.897 & 0.381 & 4.152 & 0.306 & 3.955 \\
standard & low       & 1.268 & 0.344 & 4.416 & 0.290 & 3.116 \\
large    & low       & 0.885 & 0.311 & 4.425 & 0.269 & 2.495 \\
small    & standard  & 1.905 & 0.382 & 4.192 & 0.315 & 4.184 \\
standard & standard  & 1.298 & 0.346 & 4.433 & 0.297 & 3.270 \\
large    & standard  & 0.906 & 0.314 & 4.474 & 0.277 & 2.598 \\
small    & high      & 1.929 & 0.383 & 4.222 & 0.322 & 4.401 \\
standard & high      & 1.338 & 0.349 & 4.444 & 0.303 & 3.412 \\
large    & high      & 0.938 & 0.317 & 4.505 & 0.284 & 2.700 \\
\hline
\end{tabular}
\end{table*}

\begin{table*}[h!] 
\centering
\caption{Neutral PAH flux ratios for $\log U = 3.00$.}
\label{tab:pah_ratios_neutral_U3.00}
\begin{tabular}{llccccc}
\hline\hline
Size     & Ionization & 3.3/6.2 & 6.2/7.7 & 7.7/8.6 & 8.6/11.3 & 11.3/17 \\
\hline
small    & low       & 1.771 & 0.374 & 4.314 & 0.310 & 3.952 \\
standard & low       & 1.193 & 0.337 & 4.503 & 0.296 & 3.152 \\
large    & low       & 0.808 & 0.302 & 4.491 & 0.285 & 2.638 \\
small    & standard  & 1.784 & 0.376 & 4.352 & 0.321 & 4.163 \\
standard & standard  & 1.208 & 0.339 & 4.558 & 0.307 & 3.278 \\
large    & standard  & 0.846 & 0.306 & 4.510 & 0.290 & 2.755 \\
small    & high      & 1.810 & 0.378 & 4.376 & 0.330 & 4.357 \\
standard & high      & 1.253 & 0.343 & 4.558 & 0.314 & 3.423 \\
large    & high      & 0.882 & 0.310 & 4.538 & 0.296 & 2.854 \\
\hline
\end{tabular}
\end{table*}

\begin{table*}[h!] 
\centering
\caption{Neutral PAH flux ratios for $\log U = 4.00$.}
\label{tab:pah_ratios_neutral_U4.00}
\begin{tabular}{llccccc}
\hline\hline
Size     & Ionization & 3.3/6.2 & 6.2/7.7 & 7.7/8.6 & 8.6/11.3 & 11.3/17 \\
\hline
small    & low       & 1.560 & 0.360 & 4.537 & 0.352 & 4.626 \\
standard & low       & 1.072 & 0.326 & 4.612 & 0.336 & 3.843 \\
large    & low       & 0.708 & 0.293 & 4.595 & 0.331 & 3.260 \\
small    & standard  & 1.608 & 0.365 & 4.550 & 0.357 & 4.790 \\
standard & standard  & 1.116 & 0.331 & 4.634 & 0.340 & 3.952 \\
large    & standard  & 0.754 & 0.298 & 4.602 & 0.333 & 3.363 \\
small    & high      & 1.661 & 0.369 & 4.552 & 0.362 & 4.946 \\
standard & high      & 1.152 & 0.335 & 4.657 & 0.349 & 4.056 \\
large    & high      & 0.797 & 0.303 & 4.623 & 0.335 & 3.450 \\
\hline
\end{tabular}
\end{table*}

\begin{table*}[h!] 
\centering
\caption{Neutral PAH flux ratios for $\log U = 5.00$.}
\label{tab:pah_ratios_neutral_U5.00}
\begin{tabular}{llccccc}
\hline\hline
Size     & Ionization & 3.3/6.2 & 6.2/7.7 & 7.7/8.6 & 8.6/11.3 & 11.3/17 \\
\hline
small    & low       & 2.385 & 0.396 & 3.196 & 0.352 & 6.982 \\
standard & low       & 1.528 & 0.367 & 3.562 & 0.359 & 6.062 \\
large    & low       & 1.103 & 0.347 & 3.458 & 0.355 & 5.788 \\
small    & standard  & 2.940 & 0.406 & 2.708 & 0.344 & 7.628 \\
standard & standard  & 2.104 & 0.381 & 2.727 & 0.343 & 6.865 \\
large    & standard  & 1.501 & 0.358 & 2.695 & 0.342 & 6.429 \\
small    & high      & 3.034 & 0.410 & 2.721 & 0.344 & 7.715 \\
standard & high      & 1.250 & 0.369 & 4.350 & 0.414 & 5.461 \\
large    & high      & 1.576 & 0.362 & 2.717 & 0.342 & 6.471 \\
\hline
\end{tabular}
\end{table*}

\begin{table*}[h!] 
\centering
\caption{Neutral PAH flux ratios for $\log U = 6.00$.}
\label{tab:pah_ratios_neutral_U6.00}
\begin{tabular}{llccccc}
\hline\hline
Size     & Ionization & 3.3/6.2 & 6.2/7.7 & 7.7/8.6 & 8.6/11.3 & 11.3/17 \\
\hline
small    & low       & 2.594 & 0.467 & 4.823 & 0.411 & 9.169 \\
standard & low       & 1.998 & 0.464 & 5.393 & 0.417 & 8.301 \\
large    & low       & 1.867 & 0.446 & 5.340 & 0.409 & 8.324 \\
small    & standard  & 3.454 & 0.467 & 3.672 & 0.406 & 10.170 \\
standard & standard  & 2.033 & 0.465 & 5.462 & 0.419 & 8.231 \\
large    & standard  & 1.791 & 0.454 & 5.511 & 0.417 & 8.079 \\
small    & high      & 3.486 & 0.469 & 3.664 & 0.408 & 10.242 \\
standard & high      & 2.748 & 0.457 & 4.108 & 0.410 & 9.412 \\
large    & high      & 2.217 & 0.447 & 4.533 & 0.413 & 8.902 \\
\hline
\end{tabular}
\end{table*}

\begin{table*}[h!] 
\centering
\caption{Ionized PAH flux ratios for $\log U = 0.00$.}
\label{tab:pah_ratios_ionized_U0.00}
\begin{tabular}{llccccc}
\hline\hline
Size     & Ionization & 3.3/6.2 & 6.2/7.7 & 7.7/8.6 & 8.6/11.3 & 11.3/17 \\
\hline
small    & low       & 0.041 & 0.302 & 4.760 & 2.582 & 1.782 \\
standard & low       & 0.024 & 0.304 & 4.067 & 2.468 & 1.669 \\
large    & low       & 0.019 & 0.269 & 3.228 & 2.318 & 1.553 \\
small    & standard  & 0.072 & 0.323 & 3.056 & 2.131 & 2.102 \\
standard & standard  & 0.026 & 0.286 & 4.232 & 2.638 & 1.860 \\
large    & standard  & 0.020 & 0.268 & 3.368 & 2.320 & 1.592 \\
small    & high      & 0.047 & 0.322 & 4.603 & 2.706 & 2.226 \\
standard & high      & 0.031 & 0.284 & 4.345 & 2.395 & 1.937 \\
large    & high      & 0.021 & 0.266 & 3.579 & 2.313 & 1.635 \\
\hline
\end{tabular}
\end{table*}

\begin{table*}[h!] 
\centering
\caption{Ionized PAH flux ratios for $\log U = 1.00$.}
\label{tab:pah_ratios_ionized_U1.00}
\begin{tabular}{llccccc}
\hline\hline
Size     & Ionization & 3.3/6.2 & 6.2/7.7 & 7.7/8.6 & 8.6/11.3 & 11.3/17 \\
\hline
small    & low       & 0.034 & 0.307 & 5.262 & 2.671 & 1.147 \\
standard & low       & 0.025 & 0.287 & 4.136 & 2.555 & 1.389 \\
large    & low       & 0.019 & 0.266 & 3.261 & 2.257 & 1.315 \\
small    & standard  & 0.041 & 0.302 & 5.273 & 2.696 & 1.466 \\
standard & standard  & 0.026 & 0.288 & 4.238 & 2.615 & 1.525 \\
large    & standard  & 0.020 & 0.269 & 3.391 & 2.334 & 1.389 \\
small    & high      & 0.036 & 0.312 & 5.398 & 2.841 & 1.589 \\
standard & high      & 0.027 & 0.289 & 4.335 & 2.670 & 1.648 \\
large    & high      & 0.020 & 0.271 & 3.505 & 2.355 & 1.457 \\
\hline
\end{tabular}
\end{table*}

\begin{table*}[h!] 
\centering
\caption{Ionized PAH flux ratios for $\log U = 2.00$.}
\label{tab:pah_ratios_ionized_U2.00}
\begin{tabular}{llccccc}
\hline\hline
Size     & Ionization & 3.3/6.2 & 6.2/7.7 & 7.7/8.6 & 8.6/11.3 & 11.3/17 \\
\hline
small    & low       & 0.035 & 0.304 & 5.036 & 2.227 & 0.829 \\
standard & low       & 0.026 & 0.290 & 3.886 & 2.195 & 0.980 \\
large    & low       & 0.019 & 0.267 & 3.081 & 2.096 & 1.081 \\
small    & standard  & 0.035 & 0.308 & 5.131 & 2.385 & 0.938 \\
standard & standard  & 0.026 & 0.296 & 3.995 & 2.301 & 1.155 \\
large    & standard  & 0.019 & 0.269 & 3.179 & 2.191 & 1.040 \\
small    & high      & 0.036 & 0.312 & 5.214 & 2.511 & 1.049 \\
standard & high      & 0.027 & 0.301 & 4.091 & 2.408 & 1.244 \\
large    & high      & 0.020 & 0.272 & 3.283 & 2.277 & 1.084 \\
\hline
\end{tabular}
\end{table*}

\begin{table*}[h!] 
\centering
\caption{Ionized PAH flux ratios for $\log U = 3.00$.}
\label{tab:pah_ratios_ionized_U3.00}
\begin{tabular}{llccccc}
\hline\hline
Size     & Ionization & 3.3/6.2 & 6.2/7.7 & 7.7/8.6 & 8.6/11.3 & 11.3/17 \\
\hline
small    & low       & 0.035 & 0.287 & 4.340 & 1.989 & 1.118 \\
standard & low       & 0.026 & 0.260 & 3.677 & 2.021 & 1.199 \\
large    & low       & 0.019 & 0.241 & 2.945 & 1.816 & 1.552 \\
small    & standard  & 0.036 & 0.294 & 4.564 & 2.144 & 1.212 \\
standard & standard  & 0.027 & 0.272 & 3.801 & 2.108 & 1.275 \\
large    & standard  & 0.020 & 0.245 & 3.103 & 1.894 & 1.434 \\
small    & high      & 0.037 & 0.301 & 4.735 & 2.283 & 1.312 \\
standard & high      & 0.027 & 0.282 & 4.011 & 2.186 & 1.361 \\
large    & high      & 0.021 & 0.248 & 3.248 & 1.961 & 1.418 \\
\hline
\end{tabular}
\end{table*}

\begin{table*}[h!] 
\centering
\caption{Ionized PAH flux ratios for $\log U = 4.00$.}
\label{tab:pah_ratios_ionized_U4.00}
\begin{tabular}{llccccc}
\hline\hline
Size     & Ionization & 3.3/6.2 & 6.2/7.7 & 7.7/8.6 & 8.6/11.3 & 11.3/17 \\
\hline
small    & low       & 0.034 & 0.270 & 3.786 & 2.170 & 2.122 \\
standard & low       & 0.025 & 0.245 & 3.236 & 2.197 & 1.932 \\
large    & low       & 0.020 & 0.244 & 2.347 & 1.901 & 2.091 \\
small    & standard  & 0.035 & 0.278 & 4.114 & 2.292 & 2.206 \\
standard & standard  & 0.026 & 0.255 & 3.539 & 2.218 & 1.987 \\
large    & standard  & 0.018 & 0.247 & 2.690 & 2.102 & 1.959 \\
small    & high      & 0.036 & 0.286 & 4.361 & 2.408 & 2.301 \\
standard & high      & 0.027 & 0.263 & 3.755 & 2.289 & 2.065 \\
large    & high      & 0.019 & 0.243 & 2.952 & 2.238 & 1.934 \\
\hline
\end{tabular}
\end{table*}

\begin{table*}[h!] 
\centering
\caption{Ionized PAH flux ratios for $\log U = 5.00$.}
\label{tab:pah_ratios_ionized_U5.00}
\begin{tabular}{llccccc}
\hline\hline
Size     & Ionization & 3.3/6.2 & 6.2/7.7 & 7.7/8.6 & 8.6/11.3 & 11.3/17 \\
\hline
small    & low       & 0.029 & 0.274 & 4.115 & 2.785 & 3.274 \\
standard & low       & 0.044 & 0.277 & 2.099 & 2.025 & 4.760 \\
large    & low       & 0.029 & 0.263 & 1.964 & 1.948 & 4.444 \\
small    & standard  & 0.073 & 0.292 & 2.304 & 2.210 & 5.658 \\
standard & standard  & 0.021 & 0.272 & 3.992 & 2.789 & 3.153 \\
large    & standard  & 0.032 & 0.269 & 2.013 & 2.001 & 4.543 \\
small    & high      & 0.077 & 0.302 & 2.364 & 2.295 & 5.903 \\
standard & high      & 0.053 & 0.289 & 2.218 & 2.170 & 5.022 \\
large    & high      & 0.015 & 0.267 & 3.588 & 2.793 & 3.026 \\
\hline
\end{tabular}
\end{table*}

\begin{table*}[h!] 
\centering
\caption{Ionized PAH flux ratios for $\log U = 6.00$.}
\label{tab:pah_ratios_ionized_U6.00}
\begin{tabular}{llccccc}
\hline\hline
Size     & Ionization & 3.3/6.2 & 6.2/7.7 & 7.7/8.6 & 8.6/11.3 & 11.3/17 \\
\hline
small    & low       & 0.026 & 0.318 & 6.767 & 3.569 & 5.381 \\
standard & low       & 0.019 & 0.309 & 6.874 & 3.548 & 5.122 \\
large    & low       & 0.014 & 0.299 & 6.872 & 3.513 & 5.078 \\
small    & standard  & 0.032 & 0.329 & 6.236 & 3.551 & 6.177 \\
standard & standard  & 0.021 & 0.319 & 6.761 & 3.598 & 5.203 \\
large    & standard  & 0.015 & 0.308 & 6.858 & 3.576 & 4.991 \\
small    & high      & 0.033 & 0.334 & 6.374 & 3.609 & 5.925 \\
standard & high      & 0.023 & 0.326 & 6.759 & 3.624 & 5.289 \\
large    & high      & 0.017 & 0.313 & 6.873 & 3.585 & 5.125 \\
\hline
\end{tabular}
\end{table*}

\end{appendix}

\end{document}